# A Review of Various Datasets for Machine Learning Algorithm-Based Intrusion Detection System: Advances and Challenges

**Sudhanshu Sekhar Tripathy[1], Dr. Bichitrananda Behera[2]**

**Abstract:** IDS aims to protect computer networks from security threats by detecting, notifying, and taking appropriate action to prevent illegal access and protect confidential information. As the globe becomes increasingly dependent on technology and automated processes, ensuring secured systems, applications, and networks has become one of the most significant problems of this era. The global web and digital technology have significantly accelerated the evolution of the modern world, necessitating the use of telecommunications and data transfer platforms. Researchers are enhancing the effectiveness of IDS by incorporating popular datasets into machine learning algorithms. IDS, equipped with machine learning classifiers, enhances security attack detection accuracy by identifying normal or abnormal network traffic. This paper explores the methods of capturing and reviewing intrusion detection systems (IDS) and evaluates the challenges existing datasets face. A deluge of research on machine learning (ML) and deep learning (DL) architecture-based intrusion detection techniques have been conducted in the past ten years on a variety of cyber security-based datasets, including KDDCUP'99, NSL-KDD, UNSW-NB15, CICIDS-2017, and CSE-CIC-IDS2018. We conducted a literature review and presented an in-depth analysis of various intrusion detection methods that use SVM, KNN, DT, LR, NB, RF, XGBOOST, Adaboost, and ANN. We have given an overview of each technique, explaining the function of the classifier mentioned above and all other algorithms used in the research. Additionally, a comprehensive analysis of each method has been provided in tabular form, emphasizing the dataset utilized, classifiers employed, assaults detected, an accurate evaluation matrix, and conclusions drawn from every technique investigated. This article provides a comprehensive overview of recent research on developing a reliable IDS using five distinct datasets for future research. This investigation was carefully analyzed and contrasted with the findings from numerous investigations.

***Keywords:*** *Intrusion Detection System, ML classifiers, Different IDS datasets, Evaluation matrix with accuracy, Detected assaults*

## 1. Introduction

The rapid growth of the information technology field in the last 10 years has made creating reliable computer networks a crucial task for IT managers. However, this task is challenging due to the numerous threats that can compromise the confidentiality, integrity, and availability of these networks, making them vulnerable to various risks. [1]. The Internet is a crucial tool in everyday life, used in commerce, education, medical sector, entertainment, and different fields. As technology advances, it becomes more common to use networks in various aspects of life. However, an attack on the network poses a risk due to its popularity.

IDS is a component of computer software that analyses an entire infrastructure or network of things for fraudulent behavior or adhering to restrictions. People now depend drastically on web access for practically all facets of our daily existence, as the web has completely transformed communication and our way of life. As a result, online privacy has emerged as one of the most important and pressing problems of our day. Escalating and further powerful digital attacks, crimes, and hacking resulted from our increasing reliance on digital infrastructure and software applications.

Numerous security solutions have been extensively explored and implemented throughout the years to defend against them, including firewalls, intrusion detection systems, cryptography, and encryption and decryption approaches. Due to its capacity to detect, track, and prevent intrusions by exploiting already present concepts and trends, intrusion detection [2] is regarded as the initial stage of protection against complicated and dynamic invasions [3].

Intrusion is the process of getting illegitimate entry to networks or services by tampering with the infrastructure and rendering it vulnerable. Information security is comprised of three core principles that include confidentiality, integrity, and availability. Integrity ensures data remains accurate and unaltered, while availability ensures that data are accessible to authorized users and confidentiality ensures to restrict unauthorized access and sharing personal data. Intrusion detection systems identify intruders but are susceptible to false alarms. Organizations must adjust IDS products post-implementation to prevent false alarms [4].

This review of literature examines various IDS computational algorithms, including Support Vector Machine (SVM), K Nearest Neighbour (KNN), Decision Tree Classifier (DT), Logistic Regression (LR), Naive Bayes Classifier (NB), Random Forest Classifier (RF), Extreme Gradient Boosting Classifier

*1 C V Raman Global University, Bhubaneswar–752054, Odisha ORCID ID : 0009-0003-5567-458X*
*2 C V Raman Global University, Bhubaneswar–752054, Odisha ORCID ID : 0000-0002-9362-7691*
**Corresponding Author Email: tripathysudhanshu6@gmail.com*

(XGBOOST), Adaboost Classifier (Adaptive Boosting), Artificial Neural Network (ANN), and Deep Neural Network (DNN) tests their performance on five different datasets.

### 1.1. Intrusion Detection System (IDS)

An intrusion implies that an individual operates on a system's service erroneously or without authorization. The main objective of intrusion is to compromise a resource's availability, confidentiality, and integrity. In reality, a malicious individual strives to gain access to data without authority and triggers damage to any illicit activity that might be operating.

Intrusion Detection Systems (IDS) are surveillance systems that monitor computer systems and network activity, detecting illicit activity and alerting network administrators to protect data. They enhance network privacy and data protection for enterprises, ensuring firewall security and preventing hackers from compromising secure connectivity. IDS also improves computer networks by monitoring various types of attacks.

An intrusion detection system (IDS) is a network infrastructure device that detects and alerts users of suspicious web traffic, potentially restricting traffic from dubious IP addresses, and is installed to detect anomalous activity, as depicted in Figure 1.

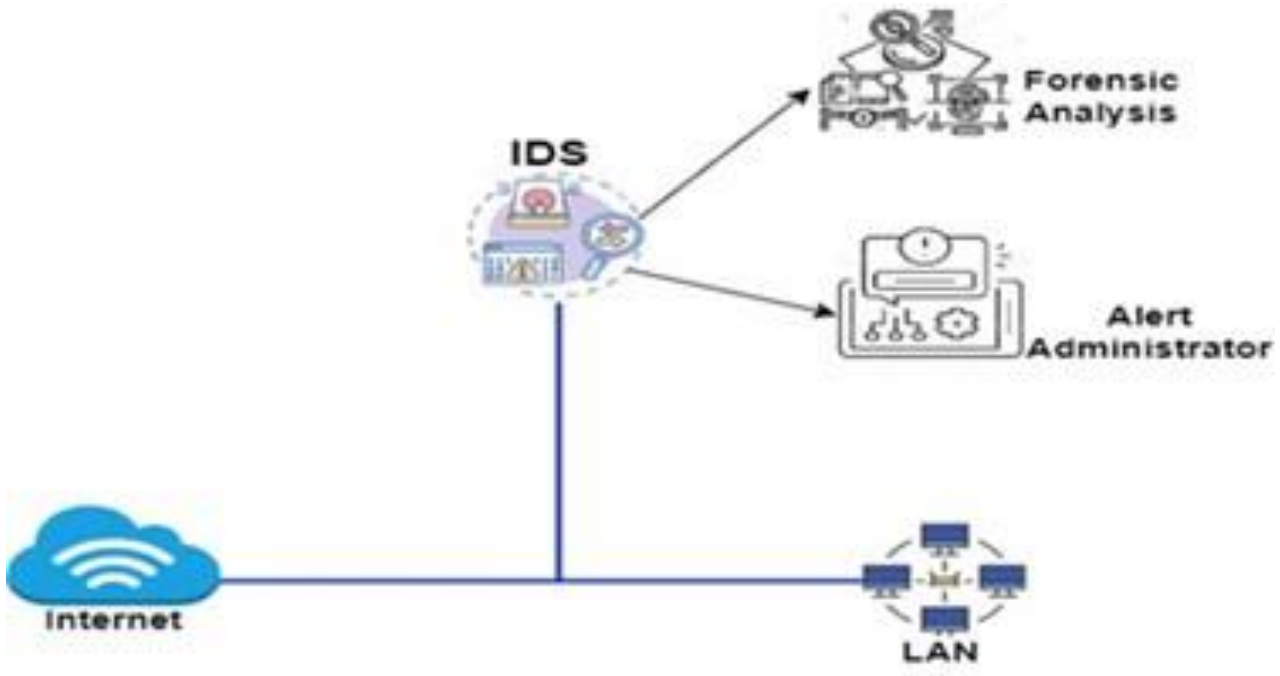

**Figure 1:** The Installation of IDS in a network infrastructure

Cybersecurity is crucial due to the internet's influence, with antivirus software, firewalls, and intrusion detection systems being key tools. The expansion of computer networks and app usage have made attacks common. HIDS and NIDS are two types of intrusion detection systems. HIDS monitors the activities on individual hosts or devices, as depicted in Figure 2, notifying administrators of suspicious activity. NIDS analyses all traffic, employing methods like packet sniffing, to detect unauthorized access. Signature detection, also known as misuse detection, relies on known pattern of attacks, and anomaly detection is a method used in NIDS, identifying deviations from normal behavior.

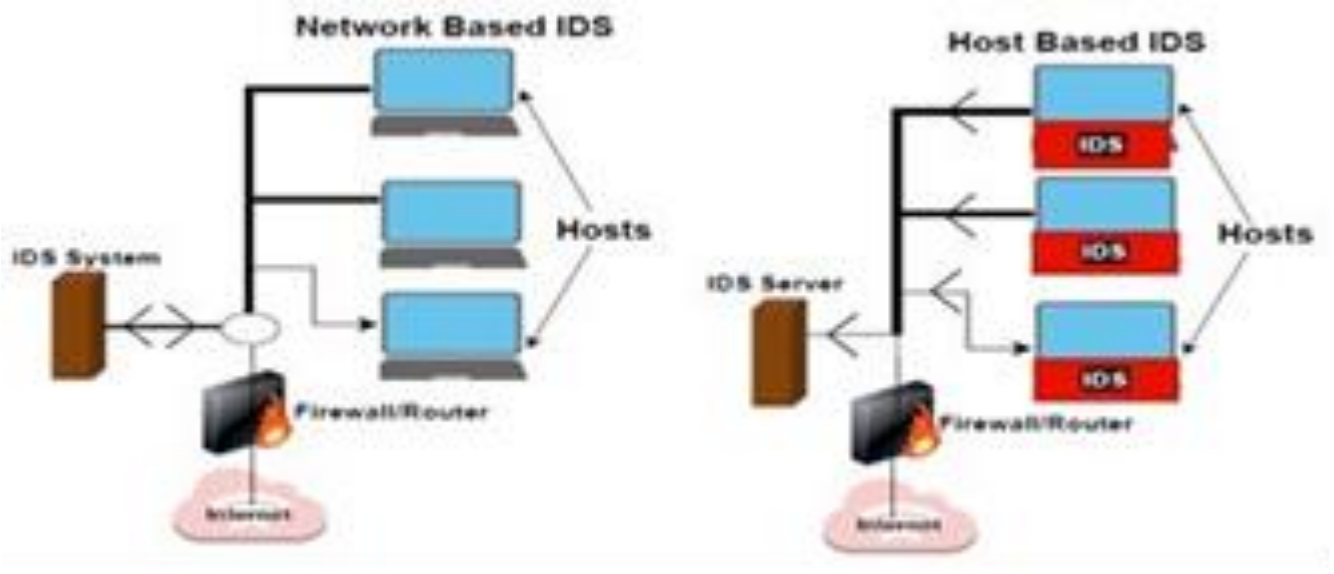

**Figure 2:** A Pictorial representation of NIDS Vs HIDS

Signature-based intrusion detection: Systems use a database of known attack signatures to identify attacks. They track network packets and generate alarms if they match. However, they can only detect registered intrusions and cannot identify new ones.

Anomaly-based intrusion detection systems, on the other hand, search for unidentified threats. It monitors network traffic and alerts system administrators of unusual behaviour, as depicted in Figures 3 and 4.

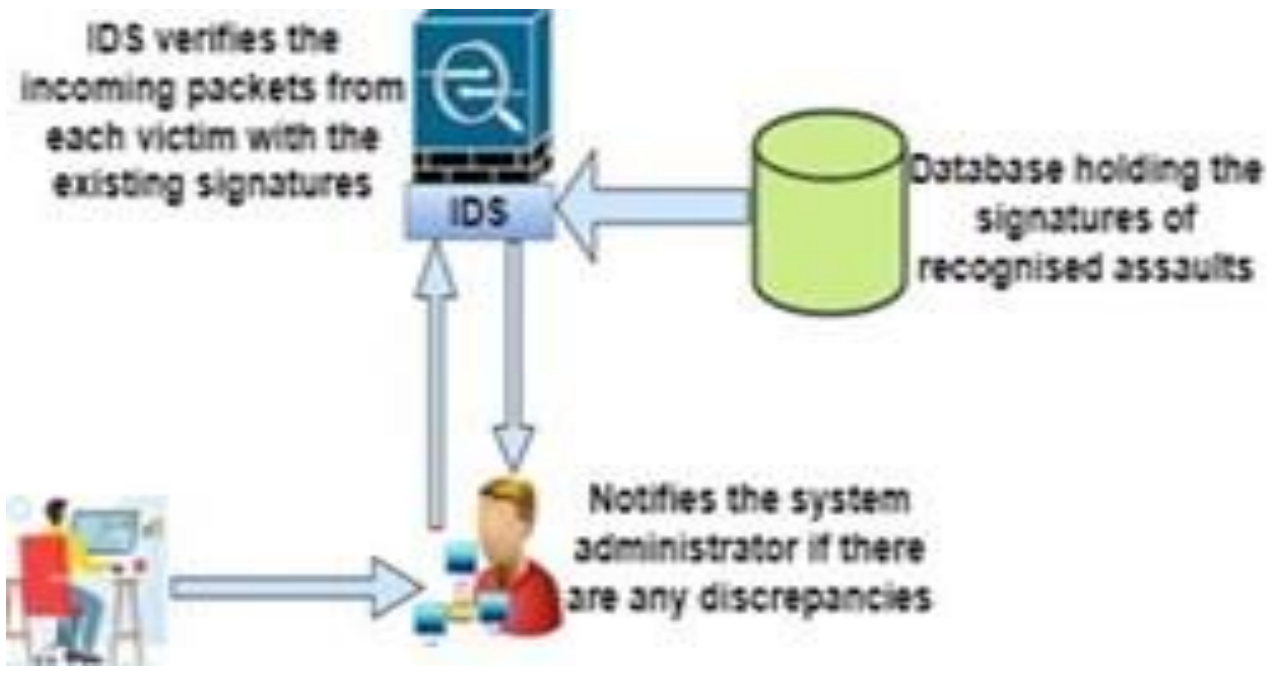

**Figure 3**: Signature-Based IDS

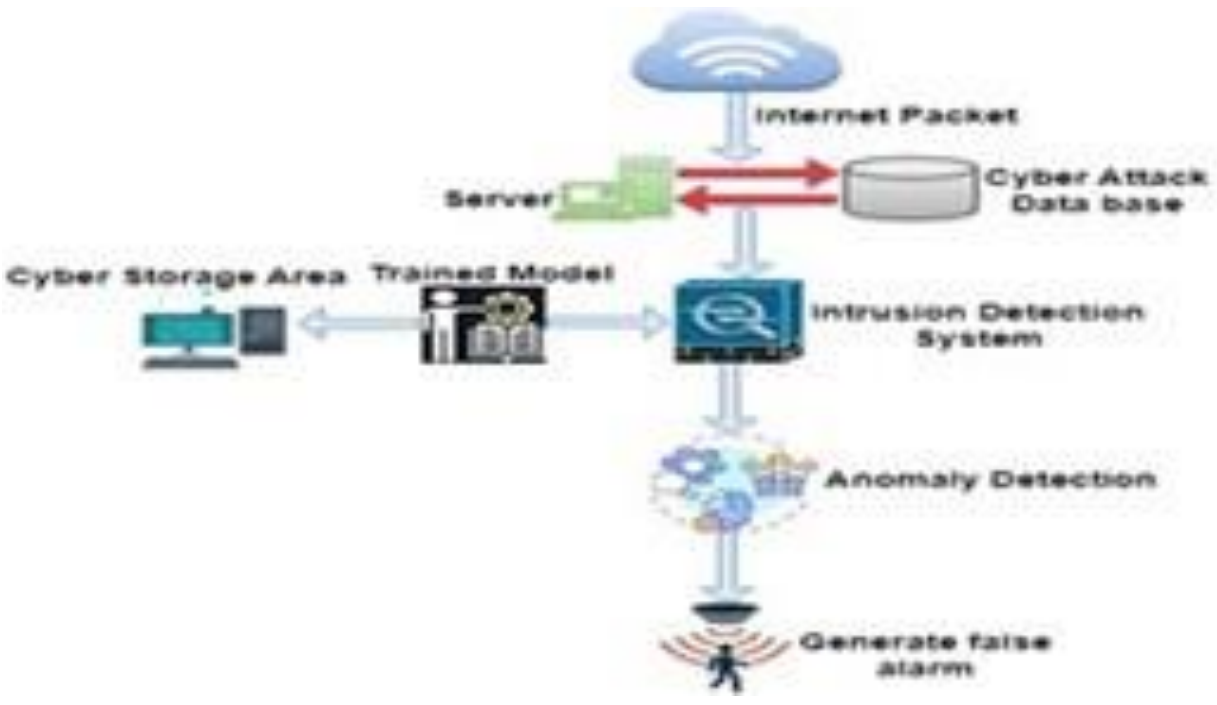

**Figure 4**: Anomaly-Based IDS

IDS sends alerts to hosts or network administrators about malicious behavior connected to a non-actively installed network switch using port mirroring technology. It monitors traffic, detects intrusions, and can be installed between network switches and firewalls, as depicted in Figure 5.

Intrusion detection involves real-time monitoring and analysis of data and networks for potential vulnerabilities and active attacks. Machine learning methods are being used for intrusion detection, but their ability to filter out false alarms is a major weakness. So, to reduce false alarms in ML-based intrusion detection, we must have quality datasets and advanced algorithms and regularly evaluate their performance.

IDS is great at spotting network attacks, but it has a big problem with a lot of false positive alerts. This can cause problems for cyber security analysts and lower the effectiveness of the system. IDSs produce fewer false positives than anomaly-based IDSs, however usage-based IDSs nonetheless frequently result in reduced detection system output because of false alarm rates. To minimise false positives, researchers are examining techniques such as machine learning, deep learning, control charts, and intelligent false alarm filters.

This paper analyses relevant literature to review the performance of current datasets from network security methods in defending computer systems from cyber-attacks, emphasizing the need for new approaches and enhanced IDS technologies.

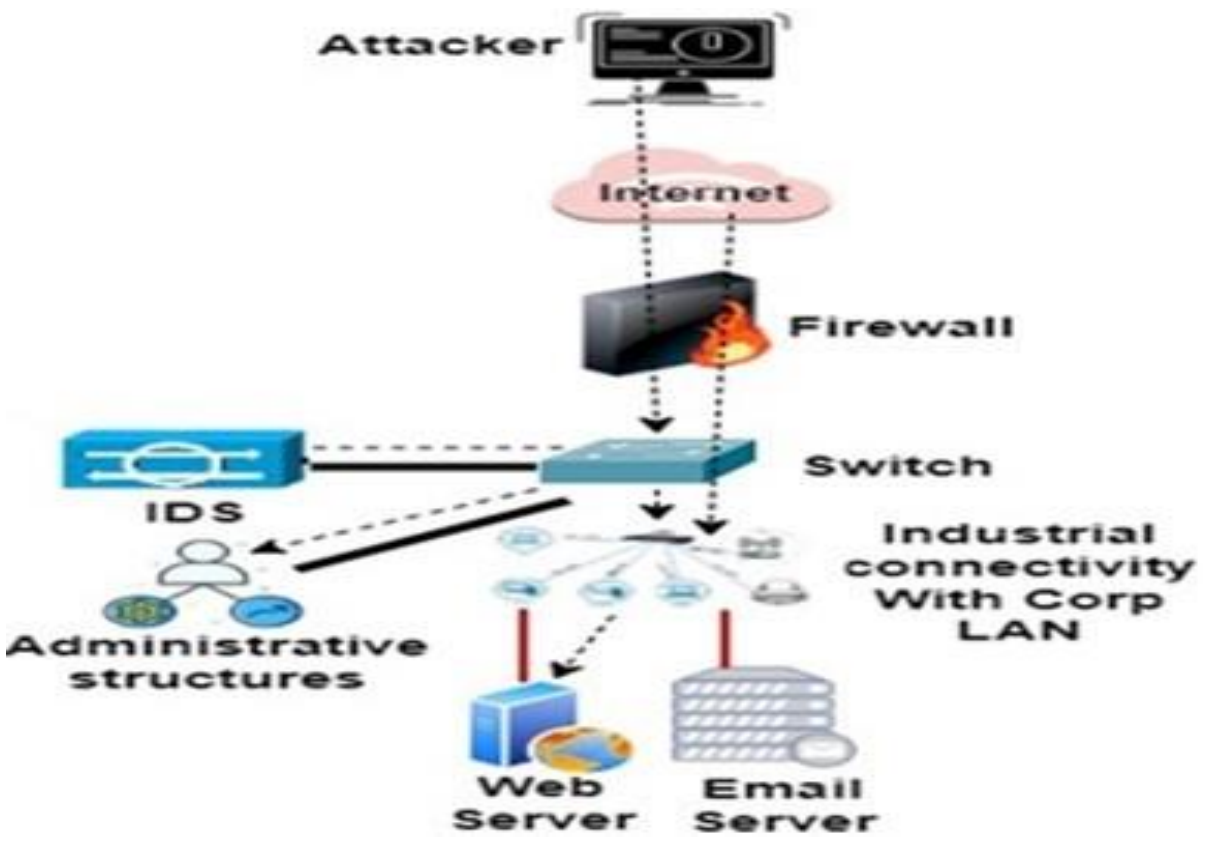

**Figure 5:** A snapshot of the Intrusion Detection System

### 1.2. Conceptual layout of the system for intrusion detection

Figure 6. depicts the intrusion detection system layout. Its principal components are as follows:

- Pre-processing phase: In this phase, packet sniffing tools like Wireshark and Capsa extract features from each packet, dividing them based on source and destination addresses. However, these tools do not determine if the packet is normal or intrusive.

- Classification: The classification phase uses data from the previous phase to determine if a packet is an attack or normal, with algorithms categorizing it into similar groups based on feature values.
    (i) Training data
    (ii) Testing data

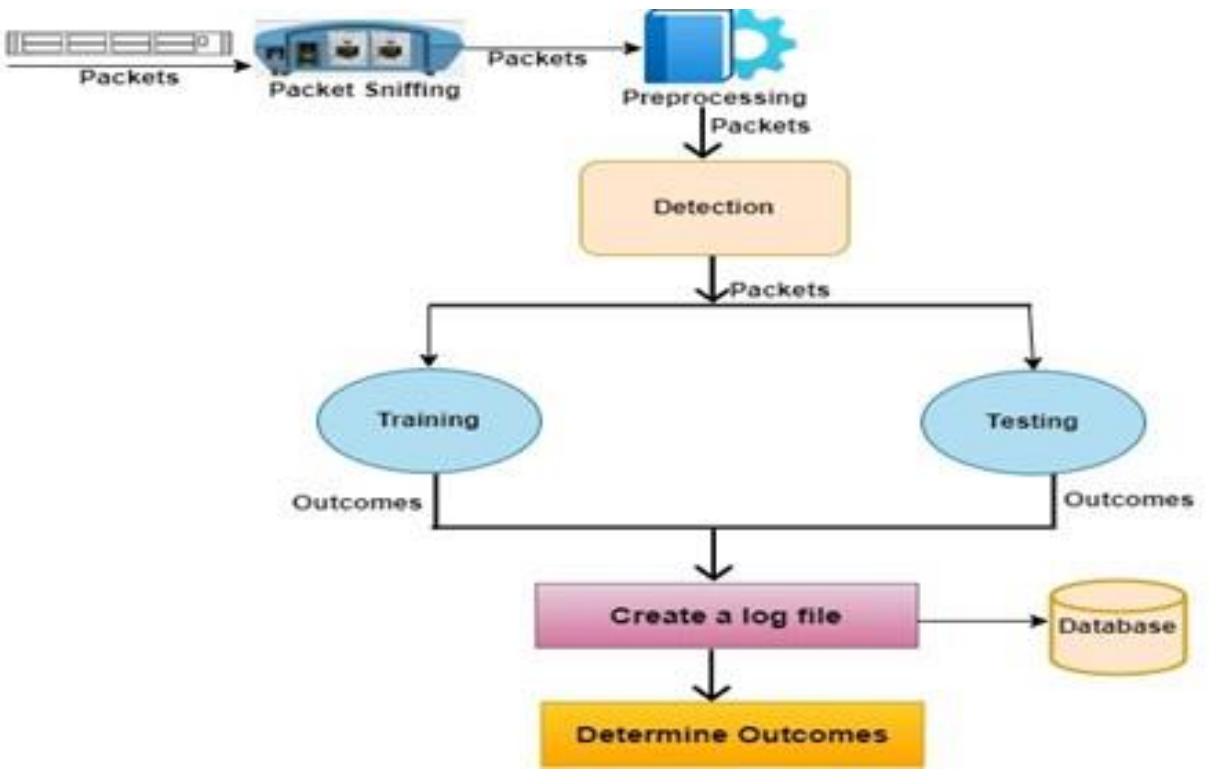

**Figure 6:** Logical layout of the system for IDS

- The training phase provides response class and packet attributes for mapping domain selection rules, which cannot be modified or replaced based on new training data during operational use, although it can be updated periodically for improved performance.

- During testing, the system uses untrained data to sample but it does not determine true answers without specifying the answer class using input packets. The test dataset contains genuine answers, which are compared to the model's predictions to assess accuracy and other performance measures.

- Reducing False Alarms: Machine learning systems require training to prevent false alarms, but human oversight and continuous learning are necessary for system accuracy and effectiveness over time.

### 1.3. The core structure of intrusion detection system (IDS)

Detecting intrusions is a strategic method that monitors network or digital setup operations, alerting potential hazards. The intrusion detection approach utilizes computational and smart techniques to identify potential intrusions through pattern identification. The intrusion detection system, a smart computer, automates the process by executing an identification model and collaborating with other IDS components. Figure 7. symbolizes the comprehensive structure of an intrusion detection system that facilitates inter-IDS component communication. The core structure of IDS includes the classification type known as the collaborative behavior of an IDS. Its foundation is each IDS's interaction and monitoring module.

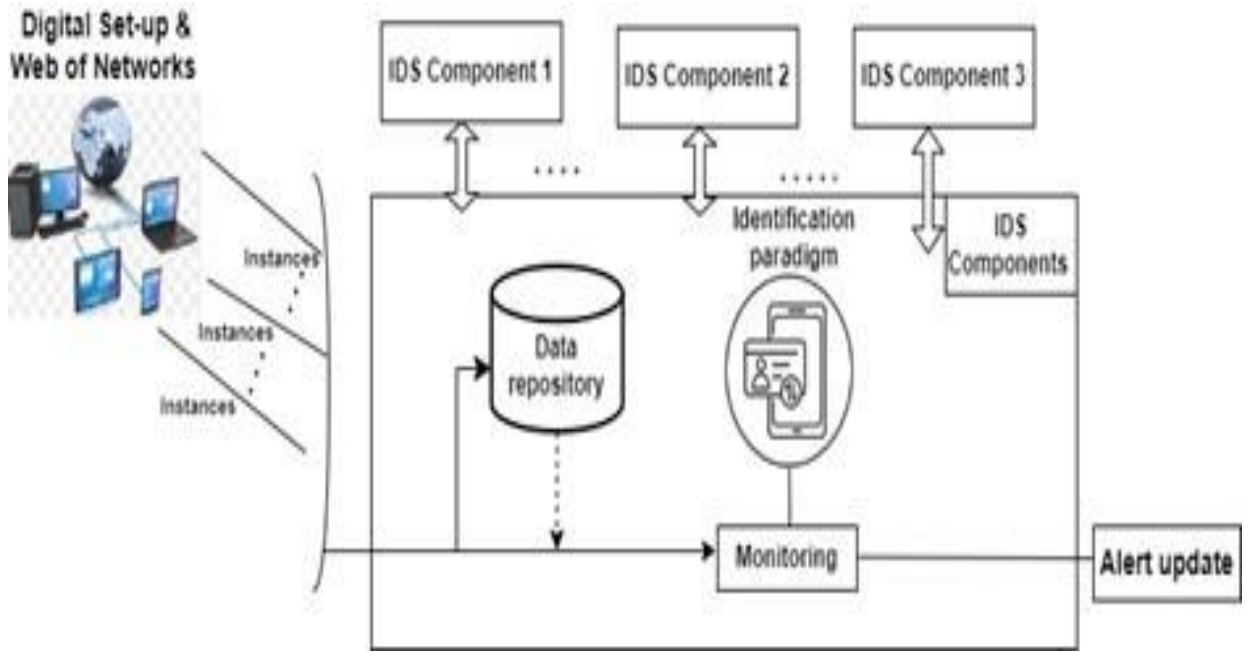

**Figure 7:** The core structure of an intrusion detection system (IDS)

### 1.4. Challenges of IDS

Intrusion detection systems are security devices that monitor network activity and computer networks to identify suspicious activity and system misuse [5]. IDSs have recently become recognized as one of the core safety mechanisms that enterprises need to implement in their IT infrastructures. When deployed with other security services, IDSs can form a layered security framework. For instance, a lot of individuals integrate IDS with antivirus and firewall software. IDSs can be employed in such a way to detect assaults that conventional security products are unable to identify.
When implementing IDS mechanisms, these particular features are crucial for an effective assault solution:

- System robustness and predictability.
- Speedy identification.
- Low false positives.
- Optimal identification accuracy.
- Reduced hardware and software prerequisites.
- Accurate intrusion location detection.
- Compatibility with other modern technology.

In conclusion, to identify assaults with outstanding accuracy and promptness, an intrusion detection system (IDS) needs to include all of the features mentioned above.

## 2. Related Work to Current Datasets

This section outlines the models of machine learning utilized in the study and presents a summary of some of the most recent datasets in the IDS domain. The IDS field faces challenges in dataset availability due to privacy and security concerns. Cybersecurity researchers have created numerous high-quality datasets for anomaly-based IDS research, which are introduced in this section.

In this part, some of the KDDCUP'99, NSL-KDD, CICIDS-2017, UNSW-NB15, and CSE-CIC-IDS2018 datasets are briefly described, along with some of key features and traits.

### 2.1. The KDDCUP'99 dataset

The KDDCUP'99 data set is a frequently used dataset for developing IDS-assessing anomaly detection techniques. In addition to the standard intrusion category (networks without any intrusions), attacks are classified into one of four types as shown in Figure 8 [6].

- Denial of Service (DoS) attacks: Effectively interpreted as flooding a network leading to service interruptions (e.g., Syn Flood).
- Remote-to-Local (R2L) attacks: Accessing a remote computer without authorization. (e.g., guessing a password)
- Probe attacks: Networks breached for surveillance. (e.g., port scanning)
- User-to-Remote (U2R) attacks: A hacker tries to log into a regular user account. (e.g., various “buffer overflow” attacks)

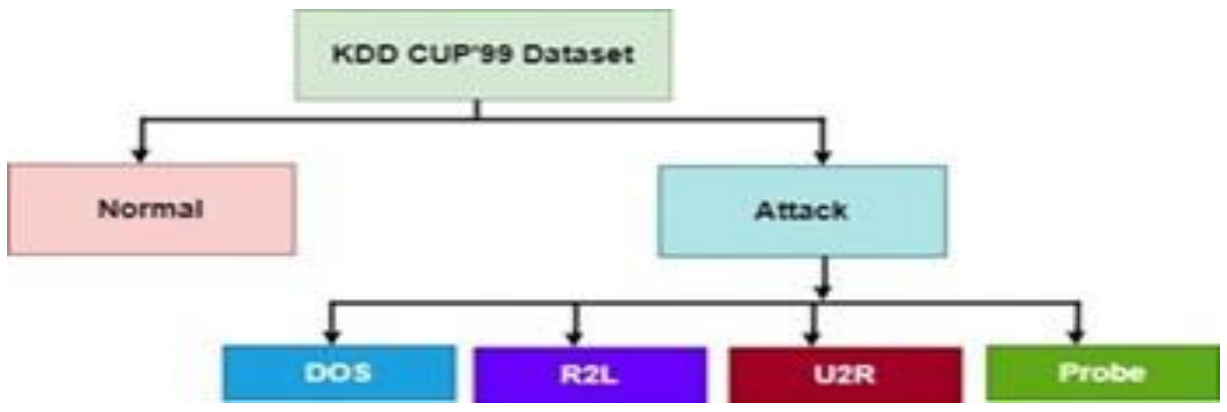

**Figure 8:** Four different types of assault categorization

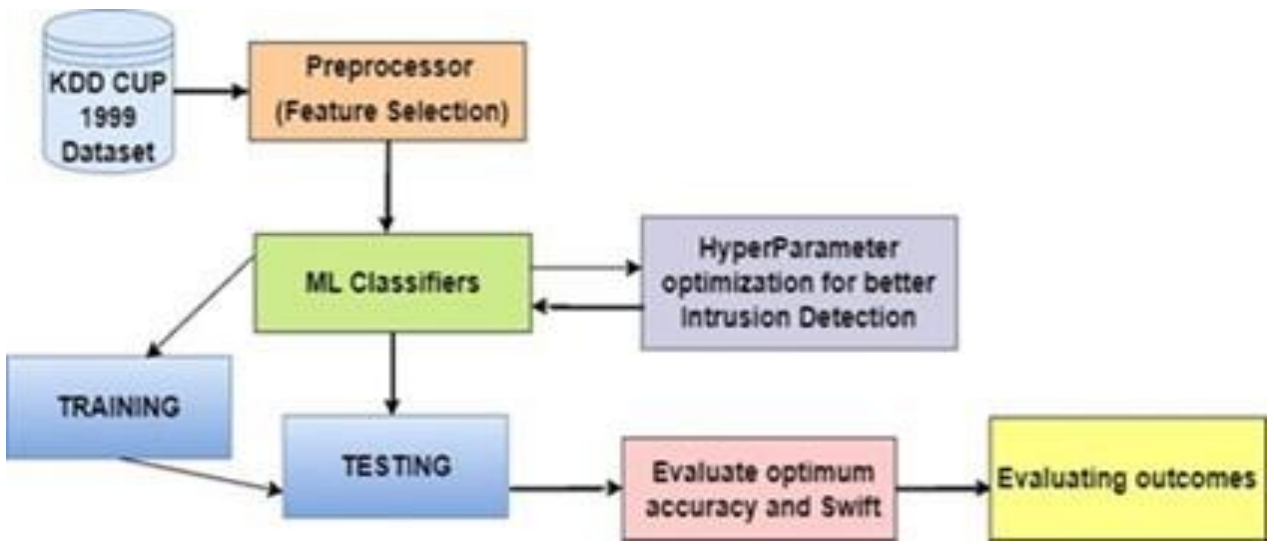

**Figure 9:** The architecture of the KDD CUP'99 assault dataset

The dataset used for the Third International Knowledge Discovery and Data Mining Tools competition aimed to develop a network intrusion model categorizing intrusions as good, bad, or normal. It consisted of 4,898,430 samples with 41 features as shown in Table 1. The initial data set included 22 attacks, categorized into four types: DOS, R2L, U2R, and probing. All attacks are classified as abnormal, requiring intrusion detection to determine the normal or abnormal mode. KDDCUP99's data set is a heterogeneous collection of 41 features with textual and numerical values, necessitating data normalization. Text attribute values are often replaced with numerical values, some of which have a large range and others with only the numbers 0 and 1, such as determining normality or assault targets in records. Then Split records into normal and abnormal cases, setting the normal record's property value to +1 and attack records to -1, scale data collection properly, and speed up computing. The KDDCUP'99 dataset, despite its diverse applications, suffers from poor performance, lengthy training, and subpar intrusion detection methods due to its high skewness, high duplicate records, and class inequalities [7].

**Table 1**. Displays the number of cases per attack category within the KDDCUP’99 datasets

| Sets | Network traffics | Authentic records | Unique records |
|---|---|---|---|
| **Training Set** | Attacks | 3,925,650 | 262,178 |
| | Normal | 972,781 | 812,814 |
| | **Total** | 4,898,431 | 1,074,992 |
| **Testing Set** | Attacks | 246,150 | 29,378 |
| | Normal | 60,591 | 47,911 |
| | **Total** | 306,741 | 77,289 |

#### 2.1.1. KDD CUP 1999 dataset

This serves as the first dataset currently being investigated as depicted in Figure 9.

- Pre-processor: Set up an approach for feature selection to eliminate anomalies and minimize the entropy of data.
- Hyper-parameters optimization: The training dataset is transformed into higher dimensions using the kernel function, ensuring linear separability by focusing on the most important hyper-parameters, like gamma. Parameter optimization is a critical aspect of improving the performance of IDS.
- ML Classifiers: The input dataset is categorized into various groups to facilitate efficient intrusion detection.
- Training Phase: The training program is trained to differentiate between legitimate and malicious data.
- Testing Phase: The suggested system is evaluated using a separate test dataset.
- Evaluate Optimum Accuracy and Swift: The categorization system's accuracy and processing speed will also be assessed.

### 2.2. The NSL-KDD dataset

The NSL-KDD dataset was introduced in 2009 to address issues with the KDDCUP'99 dataset. It is smaller, easier to maintain, and has minimal model bias due to the skewness of records. The number of sampled records is inversely proportional to the original dataset as shown in Table 2. The NSL-KDD dataset inherited most of its parent dataset's inherencies, categorizing data as normal or attack. It has 41 characteristics categorized as basic, content, or network traffic features, with 30 traffic classes in the testing set and 23 in the training set [8].

#### 2.2.1. Assaults that are incorporated in the Dataset produced by the NSL-KDD

The following section provides summaries of the four distinct categories that the dataset addresses.

- Denial of service (DoS): This kind of attack causes a server to overload with malicious requests, causing it to refuse to serve genuine traffic. (eg., SYN Flood)
- User-to-Root Attack (U2R): This is an assault where an attacker exploits a user account to gain administrative or master privileges. (eg., Buffer Overflow Attack)
- Remote to Local Attack (R2L): This attack involves an attacker transferring data via a network and deceivingly gaining local access to a machine to execute an exploit. (eg., Password

Guessing)

- Probing: This type of attack scans a network to identify its specifics and weaknesses, allowing attackers to The NSL-KDD dataset exploit these vulnerabilities for further attacks. ( eg., Port Scanning)

**Table 2**. Displays the number of cases per attack category within the NSL-KDD datasets

| Data Set Format | The overall number | | | | | |
|---|---|---|---|---|---|---|
| | **Records** | **Normal Class** | **DoS Class** | **Probe Class** | **U2R Class** | **R2L Class** |
| **KDD Train+ 20%** | 25912 | 13449 | 9234 | 2289 | 11 | 209 |
| **KDD Train+** | 125973 | 67343 | 45927 | 11656 | 52 | 995 |
| **KDD Test+** | 22544 | 9711 | 7458 | 2421 | 200 | 2754 |

The research is categorized into topics based on training and testing data, initial pre-processing, ML methods, and model outcomes. The data is initially pre-processed, normalized, and features selected as depicted in Figure 10. Pre-processing involves converting characteristics to appropriate formats, feeding machine learning algorithms, creating and training the model design, testing against test data, and achieving different anomaly detection rates using various algorithms. Performance is then compared with other models for final classification.

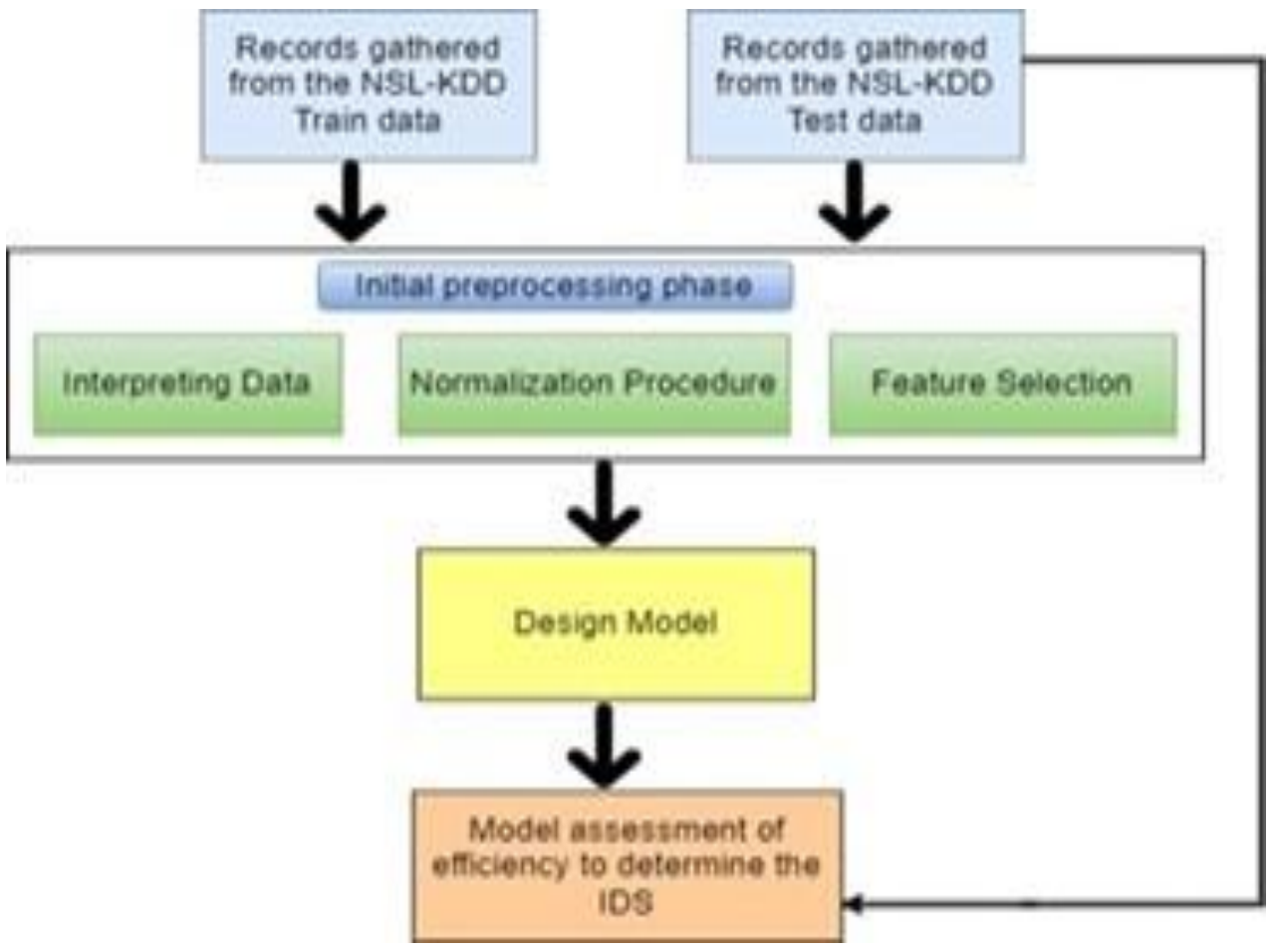

**Figure 10:** A scenario of an extensive investigative workflow employing the NSL-KDD data set for the detection of intrusions.

### 2.3. The CICIDS-2017 dataset

The CIC-IDS2017 dataset [9], created by the Canadian Institute for Cybersecurity in 2017, includes various operating systems, protocols, and attack types. It simulates 25 user actions and 2016's most frequent attacks, including port scans, infiltrations, Heartbleed attacks, botnet attacks, DoS attacks, and DDoS attacks. The dataset was publicly accessible as a CSV file on the University of New Brunswick's website [10]. However, it has a high-class imbalance problem [11], with over 70% of traffic being benign.

The CICIDS-2017 raw dataset undergoes pre-processing activities, including data creation, feature selection, and standardization. 30% is used for testing, while 70% is used for training. The trained IDS is then evaluated using the testing dataset, with outcomes evaluated based on efficiency as depicted in Figure 11.

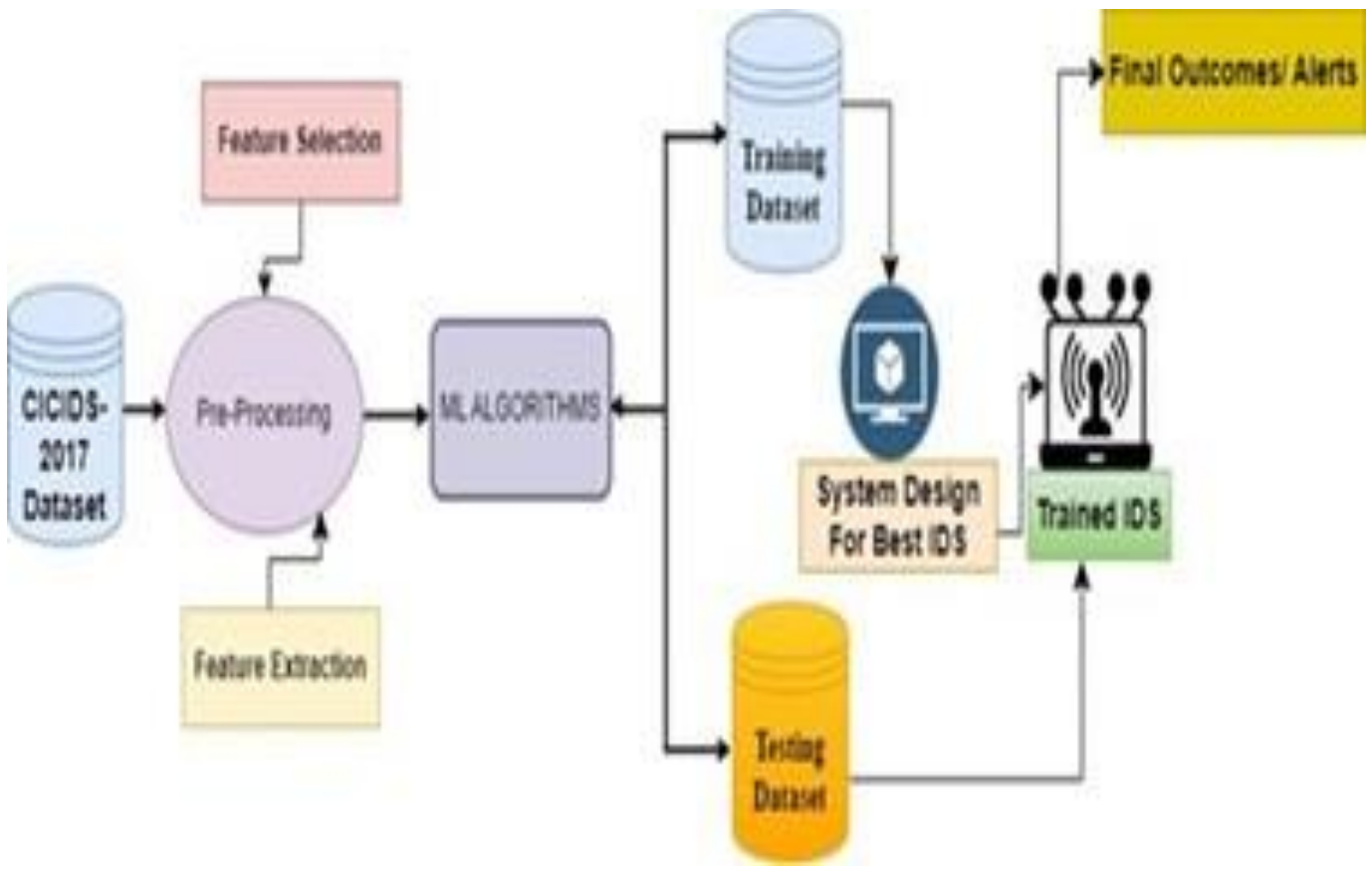

**Figure 11:** The suggested architecture of the CICIDS-2017 assault dataset

#### 2.3.1. CICIDS-2017 dataset specifications

For a total of five days, data was gathered. the event ran from Monday, July 3, 2017, at 9 a.m. to Friday, July 7, 2017, at 5 p.m. Several assaults were conducted during this time. The usual daily traffic is present on Monday. On Tuesday, Wednesday, Thursday, and Friday, respectively, the attacks took place in the morning and the afternoon. Brute Force FTP, Brute Force SSH, DoS, Heartbleed, Web Attack, Infiltration, Botnet, and DDoS are some of the attacks that have been used, as shown in Table 3.

**Table 3**. Displays the total number of Flows and Assaults within CICIDS-2017 datasets

| Day | Total Flows | No. of Assaults | Assaults Type |
|---|---|---|---|
| Monday | 529918 | 0 | Normal network activities |
| Tuesday | 445909 | 7938 | FTP-Patator |
| | | 5897 | SSH-Patator |
| Wednesday | 692703 | 5796 | DoS Slowloris |
| | | 5499 | Dos Slowhttptest |
| | | 231073 | DoS Hulk |
| | | 10293 | Dos GoldenEye |
| | | 11 | Heartbleed |
| Thursday Morning | 170366 | 1507 | Web attack-Brute Force |
| | | 652 | Web attack -XSS |
| | | 21 | Web attack-SQL Injection |
| Thursday Afternoon | 288602 | 36 | Infiltration |
| Friday Morning | 191033 | 1966 | Botnet |
| Friday Afternoon | 286467 | 158930 | Port Scan |
| Friday Afternoon 2 | 225745 | 128027 | DDoS |
| **Total** | **2830743** | **557646** | **19.70%** |

### 2.4. The UNSW-NB15 dataset

The accessible UNSW-NB15 dataset has ten classes: Normal, Fuzzers, Analysis, Backdoors, DoS, Exploits, Generic, Reconnaissance, Shellcode, and Worms, with 42 features (not including the labels). Its testing set has 82,332 records, while the training set contains 175,341. The UNSW-NB15 training and testing sets' classes are unbalanced as well. This section details the configuration of the synthetic environment and the production of UNSW-NB15, focusing on the testbed configuration and the entire process of producing UNSW-NB15 [12].

Figure 12. depicts the system's conceptual layout as it is suggested in the present research. The processing of data covers the entire first phase. Information engineering is the term employed to describe the procedure most frequently. Normalization, feature selection, and data cleaning are the three stages of information processing. Using the set of training data, the model undergoes training after the appropriate feature or set of attributes has been chosen. Next, the set of validation results is employed to confirm the accuracy of the model that was trained. Furthermore, the validated model is verified using the experimental results. Consequently, passing through the below-described method generates a customized and suitable model and results in the detection of intrusions.

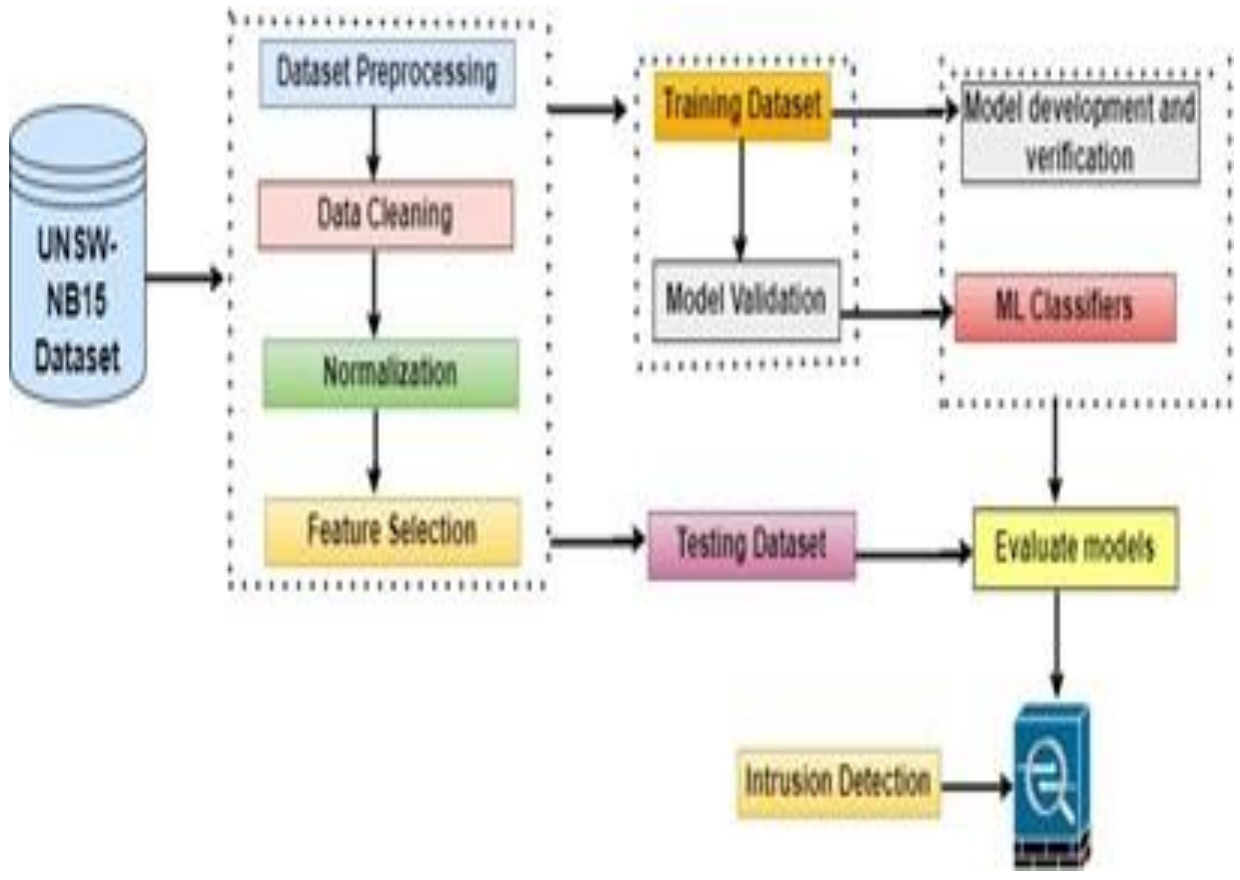

**Figure 12**: The assault dataset's recommended architecture for the UNSW-NB15 dataset

#### 2.4.1. Overview of the UNSW-NB15 Dataset

By integrating the majority of contemporary reserved assaults, the UNSW-NB15 dataset aims to emulate contemporary network settings. Table 5. lists the ten different categories of traffic that are included in the dataset: normal, fuzzing, analysis, backdoor, DOS, exploits, generic, reconnaissance, and worms. Table 4 provides an extensive overview of these categories in terms of the breakdown of training and testing sets as well as the total number of entries per assault.

The UNSW-NB15 dataset is a key tool in research for developing and testing machine learning-based intrusion detection systems, aiming to enhance detection accuracy, reduce false positive rates, and address data imbalance issues.

**Table 4.** Testing and Training for set data distribution

| Category | Testing Set | Training Set |
|---|---|---|
| Normal | 37000 | 56000 |
| Analysis | 677 | 2000 |
| Backdoor | 583 | 1746 |
| DOS | 4089 | 12264 |
| Exploits | 11132 | 33393 |
| Fuzzing | 6062 | 18184 |
| Generic | 18871 | 40000 |
| Reconnaissance | 3496 | 10491 |
| Shellcode | 378 | 1133 |
| Worms | 44 | 130 |
| **Total Records** | 82332 | 175341 |

**Table 5**. Attack categories of the UNSW-NB15

| Traffic Type | Overview |
|---|---|
| Normal | A threat-free flow of traffic |
| Fuzzing | By randomly inserting numerous data variants into a target program until one of these variations, uncovers a vulnerability, the approach automatically identifies "hackable" software flaws. |
| Analysis | Examples of this broad category include spam, port scanning, and HTML file penetration. |
| Backdoor | Illegitimate software that circumvents conventional encryption to grant remote desktop access to systems such as records and Dropbox. |
| DOS | Due to an overload of incorrect authentication attempts, the network/server could malfunction or hold up, preventing authorized users from using online services. |
| Exploits | Malware frequently contains code that exploits software flaws or security holes to spread easily and rapidly. |
| Generic | Assault by collision on the ciphers' encrypted keys. Complies with all block ciphers. |
| Reconnaissance | A set of user-friendly techniques, such as Nmap, is used to gain knowledge about a specific internet or system. |
| Shellcode | A bug introduces statements or instructions into a program that give it direct access to its registers and functions. |
| Worms | Malicious software that duplicates itself. Use excessive amounts of system memory and internet connection bandwidth. Reduces the systems' stability. |

### 2.5. The CSE-CIC-IDS2018 Dataset

This section explains the types of data that are used to implement intrusion detection. It provides genuine and rapidly changing information obtained from the platform of Amazon AWS by the Communications Security Corporation (CSE) and Canadian Institute for Cybersecurity (CIC) and depicts live traffic over the network [13]. For analyzing the detection of intrusion methods that utilize network abnormalities, it is regarded as one of the most reliable sources of input [14]. Recent attacks from 10 categories include intrusion, web, Benign, Bot, FTP-Brute Force, SSH-Brute Force, DDOS attack-HOIC, DDOS attack-LOIC-UDP, DoS attacks-GoldenEye, and DoS attacks-Slow HTTP test [15]. Table 7. provides a comprehensive breakdown of each assault class and its original collected volume. The attacked administration has 30 servers, infrastructure, 420 connections, 5 fields, and an assault architecture of 50 gadgets [16]. The CICFlowMeter-V3 program was executed to identify 80 attributes from the provided data [17]. Table 6. presents various attributes derived from the web traffic flow in Figure 13, which depicts the IDS methodology applied in the studies. More specifically, there are four steps in the method: 1) stages of datasets; 2) pre-processing; 3) training; and 4) testing.

The CSE-CIC-IDS2018 dataset is a crucial tool for developing, testing, and validating IDS models, particularly those using machine learning techniques, serving as a benchmark for evaluating detection accuracy, false positive rates, and generalization capabilities.

The CSE-CIC-IDS2018 dataset is a realistic benchmark for evaluating intrusion detection systems, encompassing 80 features from normal and malicious network traffic, and is widely used in modern IDS research.

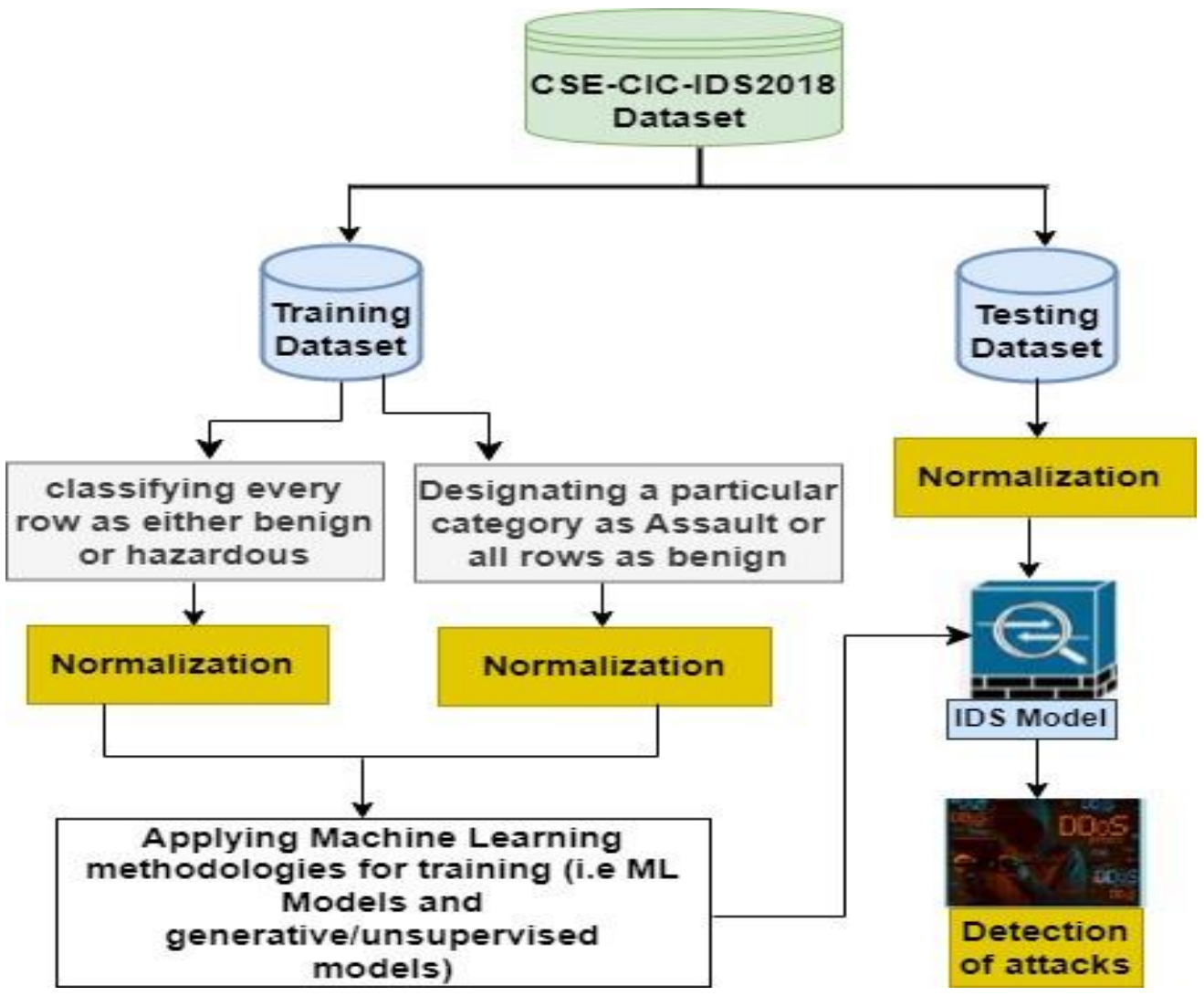

**Figure 13:** The CSE-CIC-IDS2018 attack data Sets proposed architecture

**Table 6.** Attack categories of the CSE-CIC-IDS2018 dataset discussed

| Attribute name | Description of attribute |
|---|---|
| down_up_ratio | The split between downloads as well as uploads. |
| Fl/dur | length of outflow. |
| fw/pkt/avg | The median dimension of an entirepackage traveling ahead. |
| fw/act/pkt | The percentage of communication packets currently downstream data contents for the Transmission Control Protocol, more commonly known as TCP, is at least a single Byte. |
| fw/pkt/std | The protocol packet's forward to the average difference in dimension. |
| tot/bw/pk | Total packets of information move in the reverse way. |
| tot/fw/pk | The overall amount of packets of data delivered onward. |
| Pkt/ len/var | the shortest time between payload deliveries. |
| bw/pkt/max | Maximum payload size/reverse. |
| bw/pkt/min | Minimum payload size/reverse. |
| fw/win/byt | The value of bytes transmitted in the initial session or looking forward. |
| bw/win/byt | The overall amount of bytes transmitted in the initial session or reversed. |
| bw/hdr/len | Total bytes employed in headers or backward. |
| Fw/hdr/len | Total bytes employed in headers as well as forwarding. |

**Table 7.** Illustrates a volume and percentage of points of data by assault classification

| Category designation | Assault Category | The volume of data points in a category | Ratio based on the initial data (1252835 rows). |
|---|---|---|---|
| 1 | Benign | 971016 | 77.505 % |
| 2 | Infiltration | 38703 | 3.089 % |
| 3 | DoS attacks-Hulk | 37323 | 2.979 % |
| 4 | Bot | 137185 | 10.95 % |
| 5 | DDOS attack-HOIC | 57507 | 4.59 % |
| 6 | DDOS attack-LOIC-UDP | 8377 | 0.669 % |
| 7 | FTP-BruteForce | 2234 | 0.178 % |
| 8 | DoS attacks-GoldenEye | 332 | 0.026 % |
| 9 | DoS attacks-SlowHTTPTest | 103 | 0.008 % |
| 10 | SSH-Bruteforce | 55 | 0.004 % |

## 3. Literature Review

Over the last ten years, many different IDSs have been evaluated, developed, and reviewed by implementing multiple sources of information that are publicly accessible. There have been several published reviews and comparative research studies that deal with both the ways that machine learning is implemented in developing IDS and the building blocks of IDS for various applications.

A review of the research published by Verma et al. [18] reveals that there is a possibility of greater efficiency in anomaly-based detection of intrusions, specifically concerning the degree of false positives. On the dataset provided by the NSL-KDD, learning techniques for extreme gradient boosting (XGBoost) and adaptive boosting (AdaBoost) both with and without clustering methods were executed. Even with an accuracy of 84.25%, hybrid or ensemble machine learning classifiers still need to be implemented to boost efficiency.

Dutt et al. [19] developed a hybrid detection system for intrusion tactics, identifying novel assaults using suspicious activity and prevalent assaults using misuse proximity. The anomaly identification strategy improved the algorithm's accuracy to 92.65%, and as usage increased, false negatives decreased. However, the slow detection rate persists on massive scale-dimensional data. The daily trained model reduced false negative percentages to 7.35, enhancing true positive rates and decreasing false negative rates in the abuse detection system.

Perez D. et al. [20] proposed a hybrid network-based intrusion detection system (IDS) using supervised and unsupervised machine learning strategies. They integrated artificial neural networks with feature selection and K-means clustering. The study found that the best IDS performance is achieved when SVM and K-means are combined with feature selection. To minimize false positives, hybrid technology models need to be developed.

Maniriho et al. [21] study on intrusion detection used two datasets, NSL-KDD and UNSWNB-15, with a single machine-learning algorithm and an ensemble approach. The study found superior performance with incorrect classification rates of 1.19 percent and 1.62 percent, suggesting future research should focus on improving data size and dimensionality.

Ahmad Iqbal et al. [22] implemented neural networks with feedforwards and pattern identification neural networks for IDS. They used scaled conjugate gradient methods and Bayesian regularisation. Both models performed well, with the feed-forward artificial neural network achieving an accuracy of 98.0742%. Further evaluation of multiple datasets is needed.

Watson et al. [23] developed an anomaly-based invasion detection approach using the SVM algorithm to identify malicious code at the cloud services hypervisor level. The approach achieved a 90% accuracy rate in anomalous identification, suggesting that the SVM approach could be implemented for malware identification at the lowest computational cost.

In 2019, Kaja, Shaout, et al. [24] developed a two-stage intrusion detection system (IDS) that uses an unsupervised model called k-means clustering to identify suspicious behavior, followed by supervised approaches like Naïve Bayes, random forests, and decision trees to classify harmful behavior. The system achieved an accuracy of 92.74% to 99.97% on the KDD99 dataset.

Watson et al. developed [25] an anomaly-based invasion detection approach using the SVM algorithm to identify malicious code at the cloud services hypervisor level. The approach achieved an accuracy rate of 90% in anomalous identification using system-based properties, suggesting that the SVM approach could be implemented for malware identification at the lowest feasible computational cost.

Kasongo SM and Sun Y [26] developed five supervised models: Artificial Neural Network (ANN), Decision Tree (DT), K-Nearest-Neighbour (KNN), Support Vector Machine (SVM), and Logistic Regression (LR). They used Extreme Gradient Boosting (XGBoost) to reduce feature vectors from 42 to 19 and tested on the UNSW-NB 15 dataset. DT models improved accuracy in binary classification jobs with fewer features, from 88.13 to 90.85%.

The study suggests [27] an improved method for detecting intrusion attacks using machine learning algorithms, using the KDDCUP 99 dataset. The J48, J48Graft, and Random Forest algorithms outperform other methods, with a detection rate of over 96%. WEKA is used to verify the dataset's accuracy, considering parameters like ROC, F-measure, precision, and recall. Future Improvements could involve artificial intelligence and neural networks to reduce false alarm rates.

Researchers [28] have developed a machine learning-based method for detecting intrusions in computer networks. The approach uses parameters optimization, feature selection, pre-processing, and classification to identify key attributes like Support-Vector Machines (SVM), Random Tree, AdaBoost, and K-Nearest Neighbour (KNN). Tested on large datasets like NSL-KDD and CICDDOS2019, the method outperformed other algorithms and achieved high detection rates. The study highlights the importance of DR metric values above 99% for intrusion categorization.

The study [29] uses five machine learning-based computational models, including Naive Bayes, Decision Tree, K-Nearest Neighbour, Random Forest, and Support Vector Machine, along with two deep learning models, Multilayer Perceptron Model (MLP) and Long-Short Term Memory (LSTM). The NSL-KDD dataset achieved accuracy levels of 97.77% with LSTM, 96.89% with MLP, 89.6% with normalization, and 89.2% without normalization. The neural network model outperforms conventional models in detection accuracy and incursion detection. Future improvements involve reducing the imbalance ratio and average accuracy.

Researchers [30] used machine learning and meta-heuristic algorithms to improve intrusion detection performance in the NSL-KDD dataset. They applied methods like Random Forest, Classification and Regression Trees, Support Vector Machine, and Multilayer Perceptron to maximize hyper-parameter tuning. Thestudy evaluated the effectiveness using metrics like accuracy, precision, recall, F1-score, and recall. The results showed that genetic algorithms achieved 96% accuracy, highlighting the efficiency of machine learning in cybersecurity and the potential of meta-heuristic algorithms in optimizing IDS models.

The research [31] introduces a hybrid data optimization-based intrusion detection system called DO IDS, which combines feature selection and data sampling. It removes outliers, optimizes sampling ratios, and selects the best training dataset using the Random Forest classifier. The system is constructed using the ideal training dataset and features chosen during feature selection. DO IDS outperforms other algorithms in identifying anomalous behaviors, scoring 92.8% on the UNSW-NB15 intrusion detection dataset.

The study [32] evaluates advanced intrusion detection techniques and machine learning methods. It identifies the four most effective techniques for categorizing attacks: binary, multiclass, k-nearest-neighbor, and Random Forest. Binary classification has the highest accuracy, with results ranging from 0.9938 to 0.9977. Multiclass classification outperforms the k-nearest-neighbor method, with a score of 0.9983. Random Forest's binary classification achieves the highest score of 0.9977. The study also highlights the potential of machine learning to improve system accuracy and reduce false negatives. Multiclass classification yields the best results, and distinguishing between assault types can yield more useful results.

The research [33] shows that IKPDS (Indexed Partial Distance Search K-nearest Neighbor) is a fast KNN algorithm that shortens classification completion time while maintaining accuracy and error rate for various attack types. It achieved 99.6% accuracy on 12597 cases and real-class labels, indicating that feature selection techniques can increase accuracy and save calculation time for DoS and probe attacks. The proposed algorithm maintains the same classification accuracy and takes less computing time than conventional KNN and PKDS.

Researchers [34] have developed a random forest classifier model for intrusion detection systems, outperforming conventional classifiers. Tested on the NSL-KDD data set, the model showed high detection rates and low false alarm rates. The model identified four types of attacks and underwent feature selection to reduce dimensionality. The proposed approach achieved 99.67% accuracy without feature selection, outperforming the J48 classifier by 99.26%. The model demonstrated a low false alarm rate, high DR, and good accuracy.

Researchers [35] have introduced the cluster center and nearest neighbor (CANN) technique for feature representation, which adds two distances between data samples and the cluster center and closest neighbor within the same cluster. This one-dimensional feature allows a K-Nearest Neighbour classifier to detect intrusions, outperforming 99.76% accuracy or comparable to K-NN and support vector machines on the KDD-Cup 99 dataset.

The study [36] uses machine learning algorithms to determine data intrusion rates. Support vector machines (SVMs) and Artificial neural networks (ANNs) are used to identify abnormalities or authorization issues. If malicious material is found, the request is discarded. Chi-squared and correlation-based feature selection techniques are used to minimize irrelevant data. The models are tested on a pre-processed dataset to improve prediction accuracy. The SVM algorithm achieved 48% accuracy, while the ANN model achieved 97%. Using an ANN significantly improved intrusion detection accuracy.

The research [37] presents a method for creating effective IDS using random forest classification and Principal component analysis (PCA). The strategy outperforms other methods like SVM, Naïve Bayes, and Decision Trees in terms of accuracy, with an accuracy rate of 96.78%, an error rate of 0.21%, and a performance time of 3.24 minutes.

Researchers [38] propose ML algorithms for classification, including NB, RF, J48, and ZeroR, and apply Kmeans and EM clustering to the UNSW-NB15 dataset. RF and J48 algorithms yielded the best results, with 97.59% and 93.78% respectively.

Researchers [39] have developed a network intrusion detection model using the NSL-KDD dataset and compared different machine learning techniques. The model achieved high accuracy scores of 98.088%, 82.971%, 95.75%, and 81.971% when evaluated separately. However, when combined with an inference detection model, the accuracy increased to 98.554%, 66.687%, 97.605%, and 93.914%. The study found that three out of four machine learning approaches significantly improved performance when combined with the inference detection model.

Researchers [40] use recursive feature elimination to classify unnecessary features using the KDD CUP 99 dataset and four classifier models: Random forest, SVMr, AdaBoost, and LDA. Adaboost offers the highest sensitivity and specificity, with a sensitivity of 99.75% and 95.69%. This system enhances traffic identification accuracy, increases detection rate, and reduces computation cost through feature selection. Adaboost's sensitivity and specificity make it a superior model, with a significantly higher detection rate.

The research [41] explores feature selection techniques for network traffic data for intrusion detection, focusing on discrete goal values and continuous input features. A new method is developed, achieving 99.9% accuracy in distinguishing benign and DDoS signals. The study aims to improve the understanding of network traffic data and develop robust detection systems by addressing the gap between discrete target variables and continuous input properties.

Researchers [42] found that robust SVM neighbor classification improves detection accuracy in network packet sequences, removes noise, and lowers false alarm rates. The intrusion detection rate can reach 87.3% with a false alarm rate of 0 and 100% with a false alarm rate of 2.8%.

The authors [43] have developed a novel ensemble intrusion detection system that combines decision trees, random forests, extra trees, and XGBoost algorithms. The Python-based system significantly improves detection accuracy using metrics like precision, recall, and f1-score, outperforming state-of-the-art setups. The system's performance was evaluated using the CICIDS2017 dataset, and future advancements could enhance intrusion detection and assault efficacy. The ensemble method's effectiveness is demonstrated through its performance on the CICIDS2017 dataset.

Researchers [44] have developed a unique Intrusion Detection System (IDS) method using artificial intelligence, specifically machine learning, to identify anomalies in computer networks. The model uses a Support Vector Machine (SVM) classification model with two kernels and the latest UNSWNB-15 dataset for training and evaluation. The model achieved a 94% detection rate after six measures, with 94% and 93% accuracy, respectively. Multi-class categorization is proposed for independent detection of infiltration types.

Researchers [45] have developed an intrusion detection system using machine learning, radial basis function, and multi-layered perceptron (MLP) approaches. The model's parameters are optimized using backpropagation learning and the NSL KDD dataset. Performance metrics like accuracy, sensitivity, specificity, and confusion matrix are assessed. The study found that PCA-derived features have lower false alarm rates and higher detection and accuracy rates, while RBFIDS-based intrusion detection systems offer greater accuracy than MLP-based systems, making them effective in real-world scenarios.

Researchers [46] used the UNSW-NB15 dataset to train machine learning classifiers like K-Nearest Neighbours, Naïve Bayes, Random Forest, SGD, Logistic Regression, and Random Forest. They used a taxonomy considering both eager and lazy learners and Chi-Square to eliminate redundant features. The study evaluated the effectiveness of various classifiers for intrusion detection using the UNSWNB15 dataset. The RF classifier outperformed other classifiers with a 99.57% accuracy rate, with some features alone at 99.64%.

The researcher [47] employs a wrapper method with logistic regression and a genetic algorithm to identify the optimal feature subset for network intrusion detection systems. They use the KDD99 and UNSW-NB15 datasets and three decision tree classifiers to evaluate the effectiveness of the chosen feature subsets. The KDD99 dataset demonstrated high classification accuracy with 99.90%, 99.81% DR, and 0.105% FAR, while the UNSW-NB15 dataset had the lowest FAR (6.39%) and acceptable accuracy.

The researchers [48] developed a deep-learning model using the NSL-KDD dataset to identify intruder patterns. The model achieved a 90% accuracy rate in identifying harmful network patterns and a 99.94% accuracy rate in user-to-root attacks. The F-measure showed a 99.7% accuracy rate, and the precision and recall of U2R were also 99.7%. The study used a random forest classifier for high-precision attacks.

Researchers [49] have studied intrusion detection techniques using support vector machines (SVMs) using the NSL-KDD dataset. The study found that linear SVM, quadratic SVM, fine Gaussian SVM, and medium Gaussian SVM had a total detection accuracy of 96.1%, 98.6%, 98.7%, and 98.5%, respectively, with an overall inaccuracy of 3.9%, 1.4%, 1.3%, and 1.5%. The study concluded that fine-gaussian SVM offers the best accuracy and lowest error.

The authors of [50] propose a new feature selection algorithm for Knowledge Discovery and Data (KDD) sets, focusing on dimensionality reduction in fuzzy rough sets using Maximum Dependence Maximum Significance (MDMS). The algorithm uses a modified K-Nearest Neighbourhood-based technique to categorize the data set, improve accuracy, and reduce assaults. The algorithm efficiently identifies intrusion types, providing excellent attack detection and lower false alarm rates. The modified K-NN classifier outperforms other algorithms due to its rules decision-making flexibility and fuzzy rules from the Gaussian membership function for decision-making and distance estimation, with an accuracy of detection of 98.5%.

The research [51] compares various Intrusion Detection System (IDS) algorithms, including KNN, RF, ANN, CNN, SVM, and a combination of techniques. The suggested method achieves 96.8% accuracy. The paper also examines an IDS based on Deep Q Networks (DQN), demonstrating how elegant methods can enhance IDS precision and effectiveness.

## 4. Machine Learning Classifiers for Intrusion Detection System

Intrusion detection systems (IDS) are crucial for maintaining network security by identifying attacks and unauthorized access. Machine learning classifiers significantly enhance the effectiveness and precision of IDS. Here several popular ML classifiers for intrusion detection are listed below.

### 4.1. Classification using Support Vector Machine (SVM)

Support Vector Machine (SVM) is a top learning method for binary data classification. It is primarily based on geometric principles for finding the optimal hyperplane that separates different classesas as depicted in Figure 16. SVM is also used in data privacy to identify breaches due to its high generalizability and ability to escape the dimensionality curse. As a result, SVM is increasingly used for anomalous activity detection and breach detection in data privacy.

Support Vector Machine (SVM) is a supervised learning method that classifies both linear and non-linear data by determining the optimal boundary in high-dimensional spaces as depicted in Figure 17. It is widely applied in tasks like pattern recognition and anomaly detection, including the detection of intrusions. The training process generally involves preparing the data, training the SVM model, and using it to detect anomalies or intrusions.

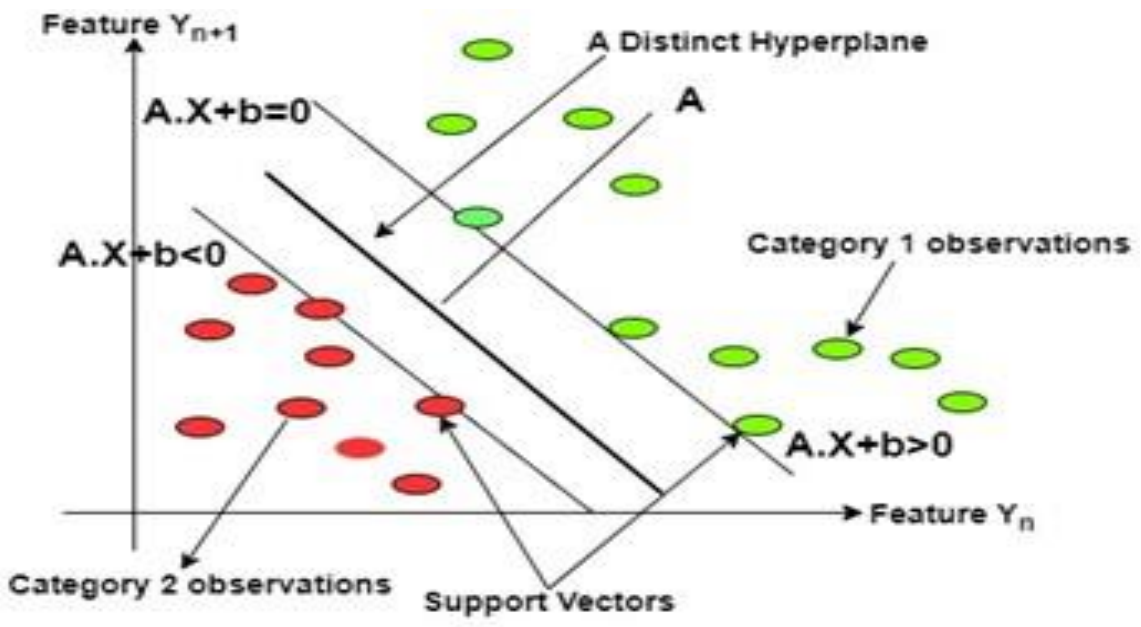

**Figure 16:** An Overview of Support Vector Machine

$f(x) = \mathbf{A.X} + \mathrm{b}$ (Linear SVM)

Or

$f(x) = \sum_{\mathbf{i=1}}^{m} a_i y_i \boldsymbol{K}(\boldsymbol{x_i}, \boldsymbol{x}) + \mathrm{b}$ (1)

(Non-linear SVM)

**The following are descriptions:**

A: The desired hyperplane in SVM maximizes the margin between classes, while the normal vector is perpendicular to this hyperplane.

X: The Support Vector Machine's Input data feed unit.

$a_i$: the Lagrange multipliers associated with each support vector.

$y_i$: Class labels.

m: The number of support vectors.

b: bias term

$K(x, xi)$: kernel functions.

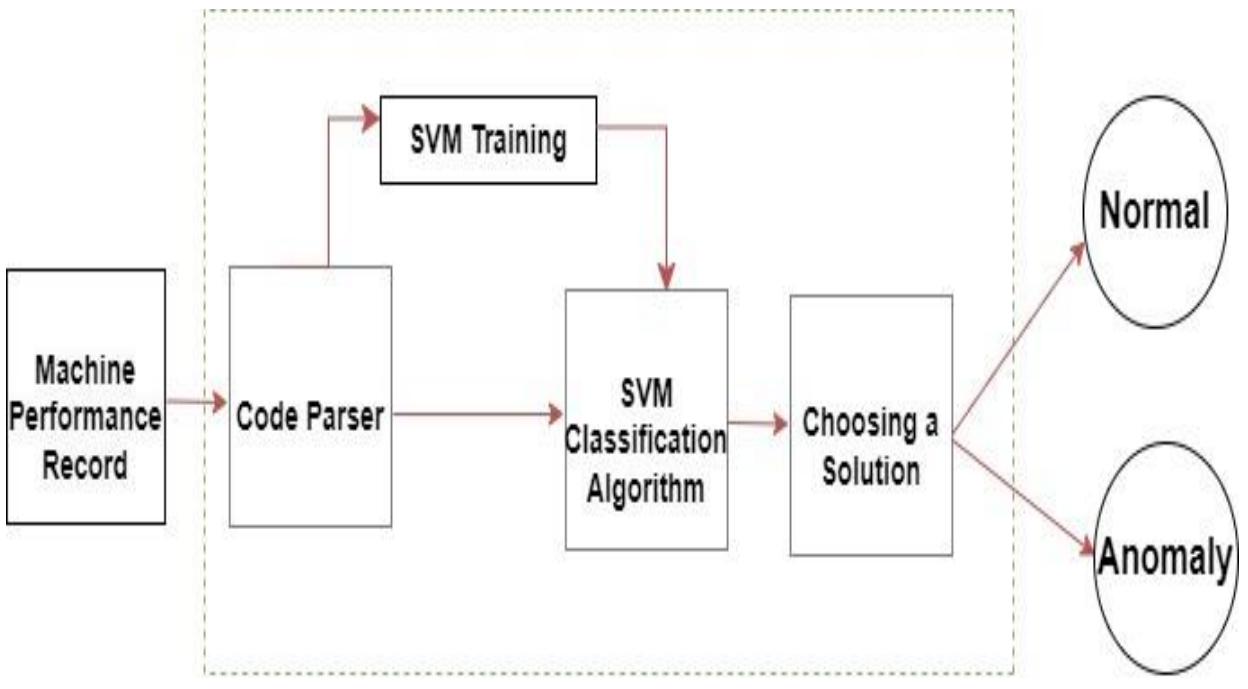

**Figure 17**: The architecture of a system for intrusion detection built on SVM

### 4.1.1. Pseudo code of SVM

**Input**: Analyse the many different types of data collected for training and testing.

**Output**: To ascertain the algorithm's accuracy.

**For** SVM, determine the ideal gamma and cost values.

**While** (Termination criterion is not completed)

**do**

**Step 1**: For every data point, carry out the SVM training phase.

**Step 2**: Execute the SVM method to evaluate the testing data point.

**Step 3**: State the SVM-based kernel as K (x, y) = $e^{-\left(\frac{||x-y||^2}{2\sigma^2}\right)}$

where x and y are objects of the attributes range in each parameter of the training set.

**Step 4**: The goal is to implement two classes:

1 – Normal

2 – Anomaly

**Step 5**: End while

**Step 6**: Return accuracy.

### 4.2. Classification using K-Nearest Neighbor (KNN)

K-nearest neighbor (KNN) is a non-parametric supervised classifier that uses distance as a Euclidean measure [52] to predict desired parameter results. It classifies program behavior as normal or intrusive, producing results only when requested. KNN analyzes K instances of training data closest to the test sample and assigns the most frequently occurring class label.

The KNN classification method categorizes new data into previously observed classes based on the majority class of its nearest neighbors. In the Figure 18, the new instance to classify is a black hexagon, with blue squares representing normal behavior and orange triangles representing abnormal behavior.

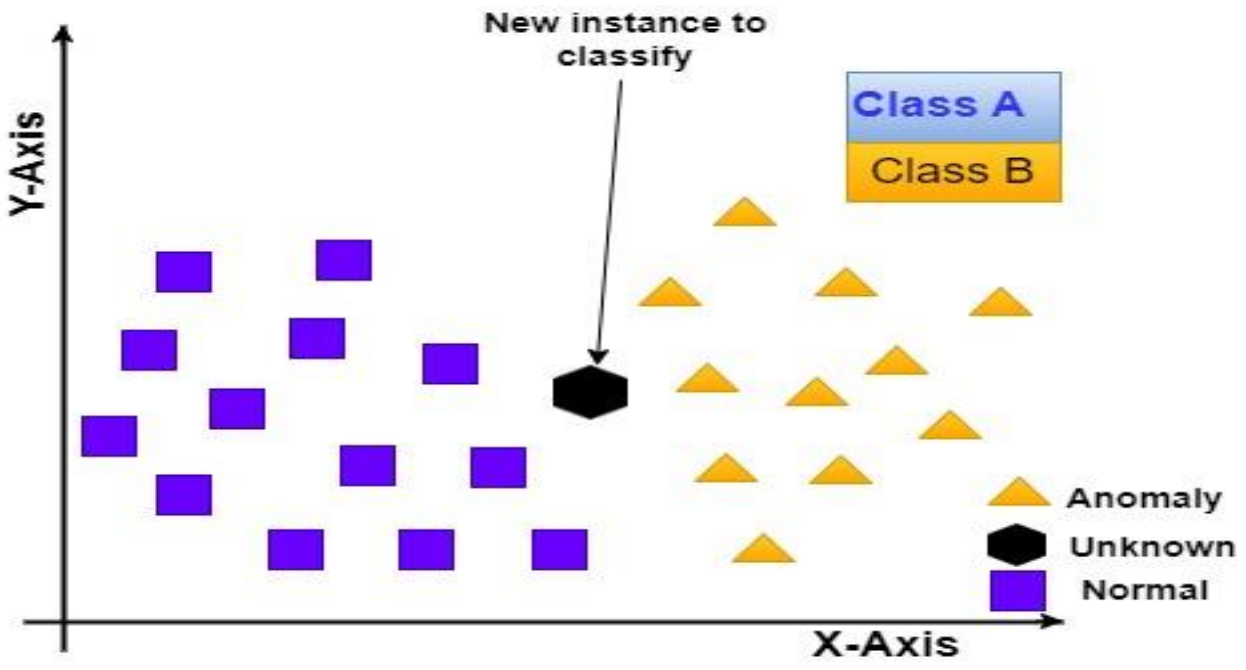

**Figure 18**: K- Nearest Neighbors (KNN) Classifier

### 4.2.1. Pseudo code of KNN

**Step 1**: Loading the training and testing datasets.

**Step 2**: Initialize k as the value of the neighbors.

**Step 3**: From step 1 to the entire number of training data points, repeat this process to obtain the predicted class.

- Calculate the distance between training and test data using Euclidean distance, Manhattan distance, Minkowski distance, Chebyshev, cosine, or other metrics, or hammering distance for categorical variables.
- Using the distance values as a basis, sort the computed distances into ascending order.
- Extract the top k rows from the array that has been sorted.
- Find the class that appears the most frequently among these rows.
- Return the predicted class.

### 4.3. Classification using Decision Tree Classifier (DT)

The decision tree algorithm is a widely used classification algorithm that uses a tree-shaped graph to classify objects based on rules applied to the tree's leaves. It is particularly effective for intrusion detection, where connections and users are classified as normal or attack types based on pre-existing data. Decision trees, which learn from training data and forecast future data, are effective for large data sets and real-time intrusion detection due to their high generalization accuracy [53]. They help create clear security protocols and can be used with minimal processing in rule-based models. Decision trees are used to check attribute values of a network traffic profile tuple, X, with unknown class labels, predicting the leaf node's class label as depicted in Figure 19.

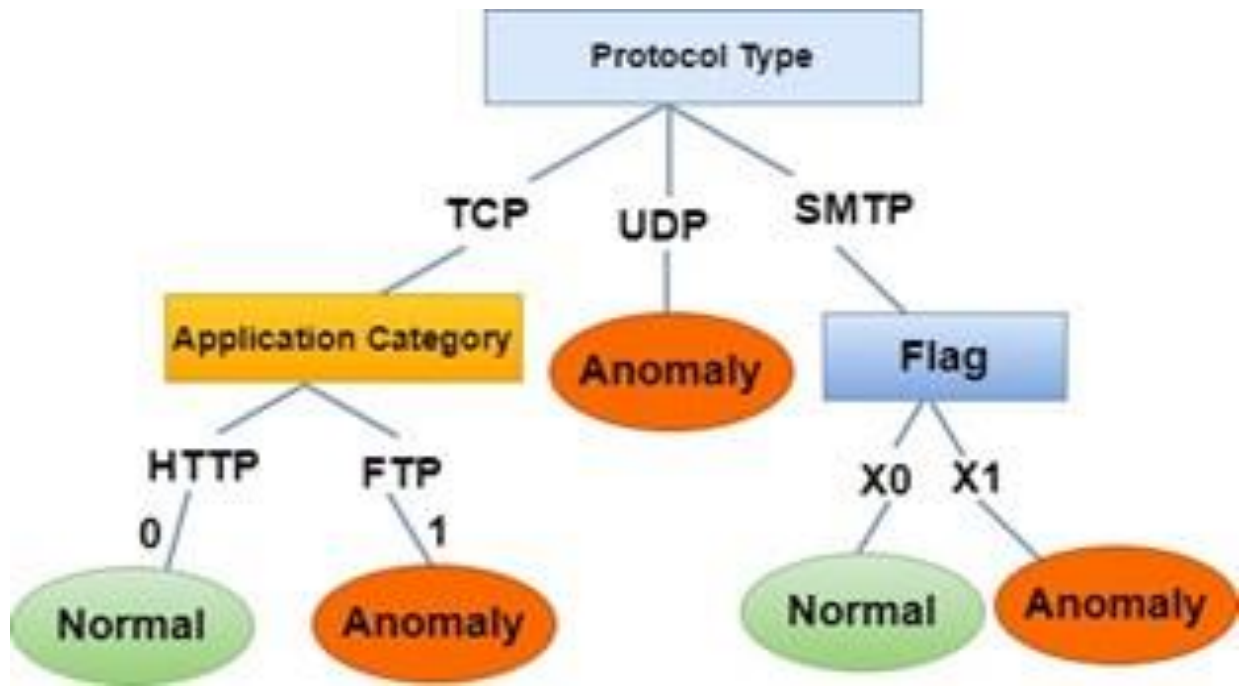

**Figure 19**: Decision Tree Classifier

#### 4.3.1. Pseudo code of DT

**Input**:
T // Decision Tree
D // Input Data set
**Output**:
M // Model Prediction
**Step 1**: Select any input feature from the data set.
**Step-2**: for each t ∈ D
**do**
**Step 3**: n = root node of T;
**Step 4**: While n not a leaf node
**do**
**Step 5**: Obtain the answer to question to on n applied t;
**Step 6**: Identify an arc from t that contains the correct answer;
**Step 7**: n = node at end of this arc;
**Step 8**: Predict t based on labeling of n;
**Step 9**: End while
**Step 10**: Return accuracy.

### 4.4. Logistic Regression Classification (LR)

Logistic regression is a machine learning algorithm used to predict binary outcomes, such as normal or anomaly, true or false, or whether an event occurs or fails. It uses a categorical dependent variable and independent factors to determine the binary outcome. Similar to linear regression, logistic regression is used for classification problems and linear regression for regression problems. The "S" curved logistic function predicts the two highest possible outcomes (0 or 1) and is calculated within a regression model. The supervised machine learning method LR is used to observe the discrete collection of classes. The logistic function uses the sigmoid function, also known as the cost function, which maps predictions to probabilities [54], allowing for the prediction of an event's probability as shown in Figure 20.

$$P\ (B = 1|A) \text{ or } P\ (B = 0|A) \qquad (2)$$

In this case, the independent variable is A, and the dependent variable is B. Logistic regression makes use of the sigmoid function.

$$F(x) = \frac{1}{1+e^{-x}} \qquad (3)$$

F(x) returns a value between 0 and 1, where e is the natural log base and x is the function's input parameter.

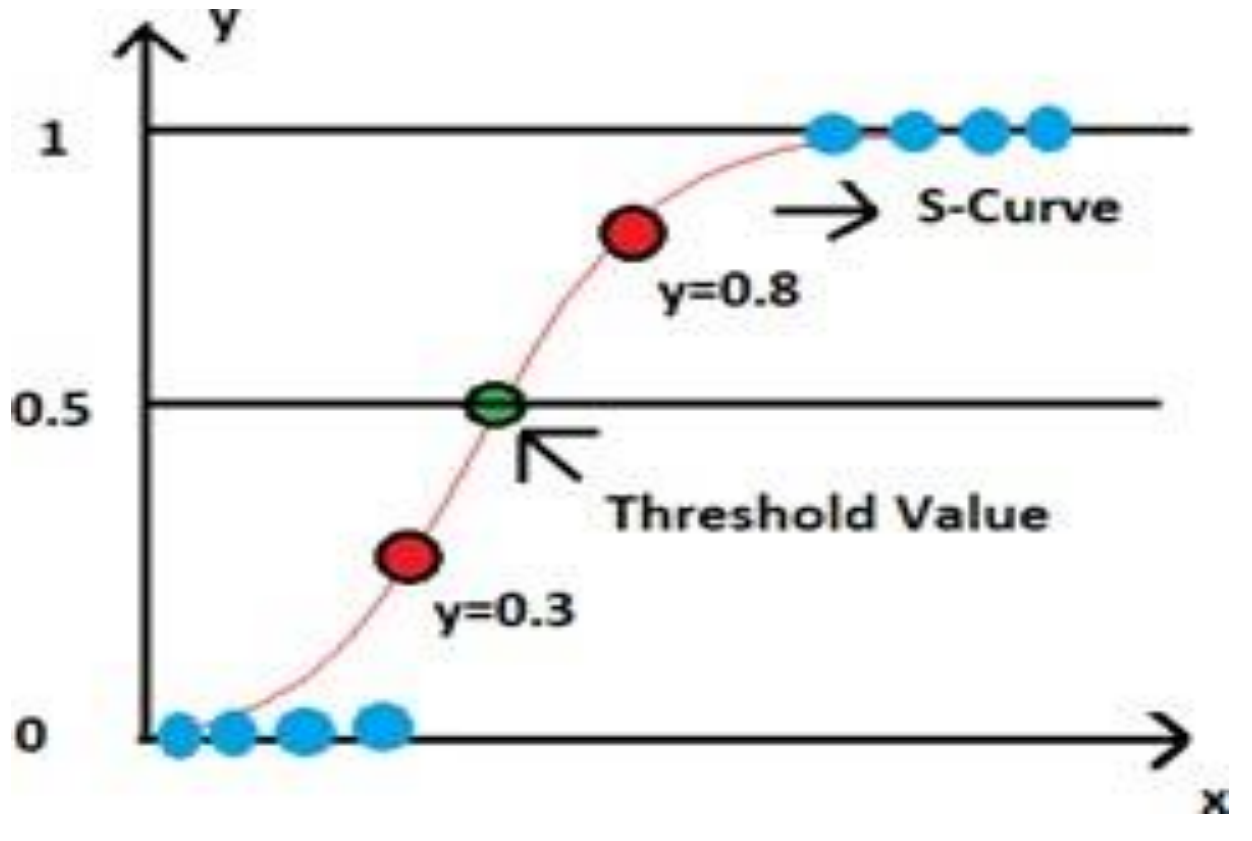

**Figure 20**: Logistic Regression (LR)

### 4.5. Naive Bayes Classifier (NB)

The Naïve Bayes model operates on Bayesian principles and relies on the assumption that features are independent of each other when conditioned on the class label. Although it is a straightforward model, it often delivers precise outcomes. Misclassifications in Naïve Bayes can result from noisy data and the model's inherent bias due to its independence assumption [55]. To reduce the effect of noise, using high-quality data is crucial. Unlike clustering methods, Naïve Bayes does not divide data into distinct groups but estimates probabilities based on the assumption that feature dependencies do not exist.

Based on the Bayes theorem, the naïve Bayes model operates. It also makes use of naive assumptions. The chance of one event occurring when the likelihood of another is known to be found using Bayes's theorem. Bayes Theorem mathematical formula:

$$P\ (S|T) = \frac{P(T|S).P(S)}{P(T)} \qquad (4)$$

Where T: The details combined with unlabelled groups.
S: The specific group is represented by the T facts statement.
P(S|T): The likelihood of an assumption S is dependent on the presence of state T.
P (S): Assumption likelihood S.
P (T|S): T likelihood depending on the state.
P (T): T likelihood.

#### 4.5.1. Pseudo code of Gaussian Naive Bayes

**Input**: Training dataset Td,
P = (p1, p2, p3, ……., $p_n$) // measurement of the evaluating testing dataset's predicted variable's value.
**Output**: A category of datasets for testing.
**Step 1**: Read the training dataset Td;
**Step 2**: Determine the predicting variables in each class's mean and standard deviation;
**Step 3**: Repeat
• Applying the Gaussian Density Equation, determine the probability of pi for each category.
• Until the estimated likelihood of every variable that predicted (p1, p2, p3, $p_n$) has been determined.
**Step 4**: Determine each class's likelihood;
**Step 5**: Obtain the most significant likelihood.

### 4.6. Random Forest Classifier (RF)

The Random Forest model, also known as Random Decision Forest, is a classification technique that uses decision trees to make decisions based on the majority's recommendations as depicted in Figure 21. It is composed of an ensemble of decision trees assembled from a bootstrap sample from a training set, with three hyper-parameters established before training. It is flexible for regression and classification issues [56].

Random Forest is a popular intrusion detection technique used to identify anomalies in large datasets. Its strength lies in its ability to handle complex, high-dimensional data, making it effective for detecting unusual patterns. However, it may face challenges with small datasets, potentially leading to overfitting or less reliable performance. Additionally, its computational demands may impact performance in certain scenarios, especially with large datasets.

### 4.6.1. Pseudo code of Random Forest

**Step 1**: From the given training data set, K data points are randomly selected by the method.
**Step 2**: The process involves creating a decision tree for each data point and extracting each tree to obtain an estimate.
**Step 3**: When building decision trees, select the number N.
**Step 4**: Go back and repeat steps 1 and 2.
**Step 5**: The algorithm will select the most highly voted-for predicted outcome as the final estimate.

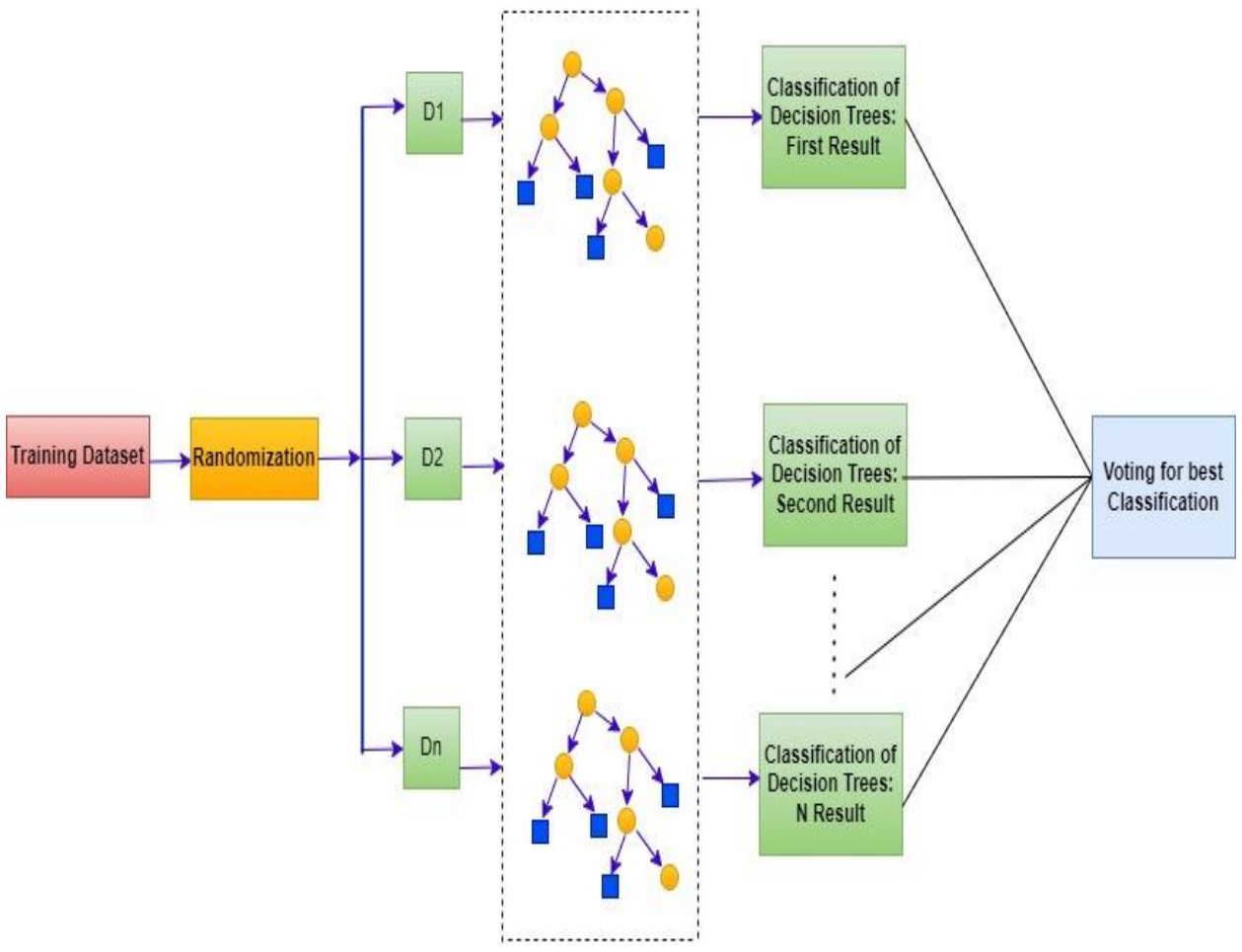

**Figure 21:** The structural framework of the Random Forest Algorithm

### 4.7. Extreme Gradient Boosting (XGBOOST) Classifier

XGBoost is a sophisticated ensemble learning technique that enhances the performance of sequential decision tree algorithms through a method known as gradient boosting. It operates within the Gradient Boosting Decision Tree (GBDT) framework and is designed to handle distributed computing efficiently. XGBoost boosts model accuracy by addressing the residuals left by previous models and incorporating both first and second-order derivatives of the error function in its optimization process. This "boosting" technique combines multiple models to address and correct errors from earlier iterations, resulting in a more robust overall model. Using decision trees as its base learners, XGBoost is scalable and applicable to a variety of tasks, including classification, regression, and ranking, making it effective for improving model performance and generating predictions as shown in Figure 22.

XGBoost is a machine learning algorithm that enhances model performance through gradient boosting, regularization to prevent overfitting, and memory optimization techniques. Its efficiency and scalability are enhanced by features like data and feature subsampling, making it ideal for handling large and complex datasets.

### 4.7.1. Pseudo code of XGBOOST

**Input**: Training and Testing Dataset
**Output**: Datasets labeled as either normal or assault, depending on the individual category designation.
**Step 1**: Evaluate and cleanse the input dataset.
**Step 2**: Implementation of the min-max approach to normalize the input dataset.
**Step 3**: Implement the XGBoost prominence of feature values while selecting attributes.
**Step 4:** Applying the selected set of attributes from Step 2, develop the ML classifier.
**Step 5**: Train the ML classifier.
**Step 6**: Apply the ML optimizer to the ML classifier that was built in steps 4 and 5.
**Step 7**: Apply the k-fold cross-validation method to evaluate the XGBoost classifier model.
**Step 8**: The final prediction will be obtained by the algorithm.

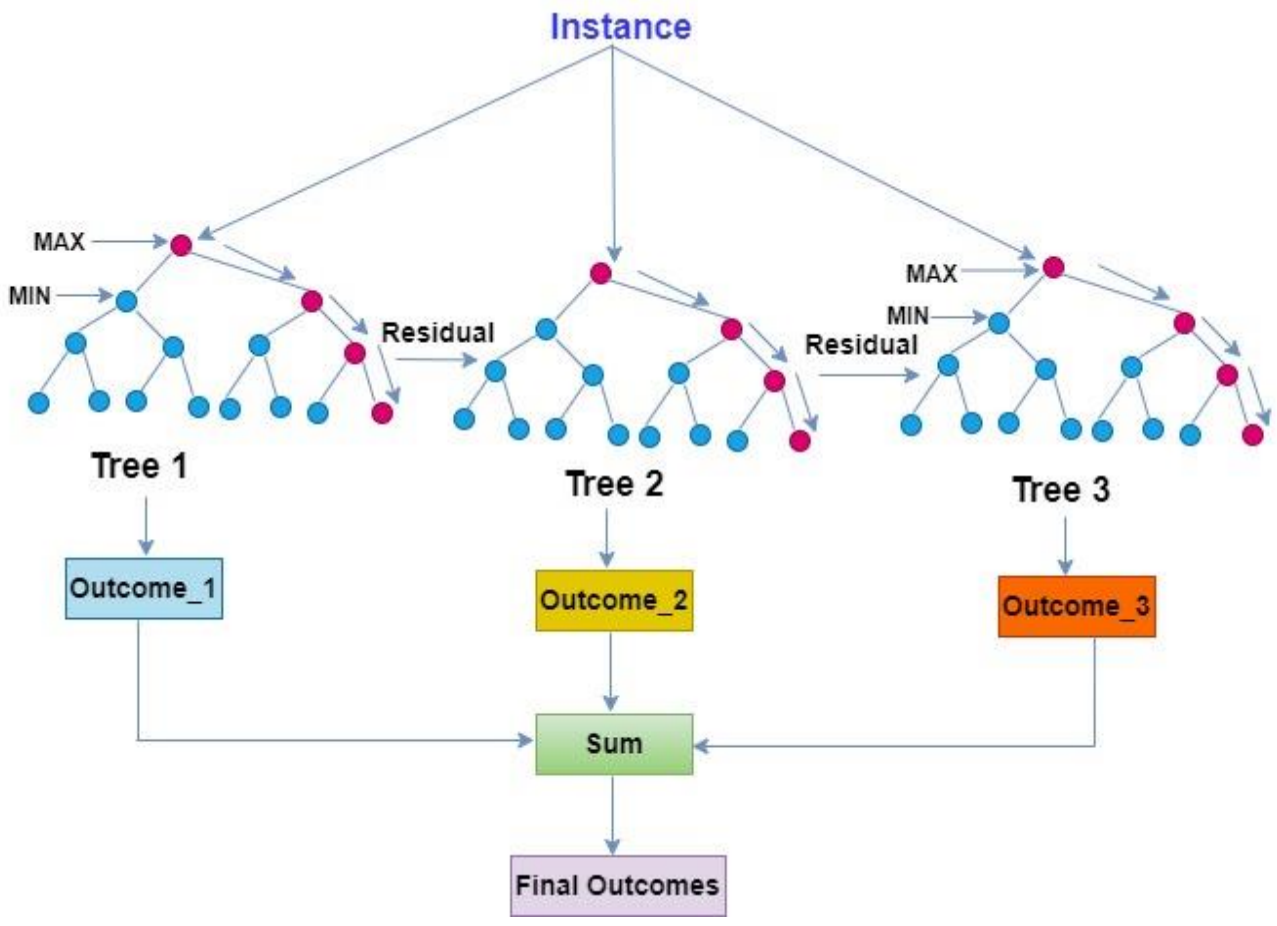

**Figure 22:** An overview of the XGBOOST System's architecture

### 4.8. Adaboost Classifier

AdaBoost is a boosting technique that builds a strong model by combining multiple weak classifiers in a sequential manner. The process involves training weak classifiers iteratively, each time adjusting the focus on misclassified instances from the previous iteration. This method enhances the performance of the ensemble by weighting these weak classifiers according to their accuracy as shown in Figure 23 . AdaBoost's approach of integrating numerous weak models helps in creating a robust classifier that performs well on a variety of datasets. This technique is valued for its ability to improve classification accuracy through iterative refinement and weighted learning.

### 4.8.1. Pseudo code of Adaboost

**Step 1**: Adaboost applies the entire training set but assigns weights to each instance.
**Step 2**: The AdaBoost machine learning model adjusts the weights of the existing instances based on their classification errors in the previous iteration.
**Step 3**: Erroneously classified observations are given a higher weight to increase their likelihood of being correctly classified in the subsequent iteration.
**Step 4**: The trained classifier's weight is determined by its accuracy, with higher weights indicating greater accuracy.
**Step 5**: The procedure is repeated until the entire training set is error-free or the maximum number of estimators is reached.
**Step 6**: The final model is a weighted combination of all the weak learners.
**Step 7**: The final method of calculation will determine the most accurate prediction.

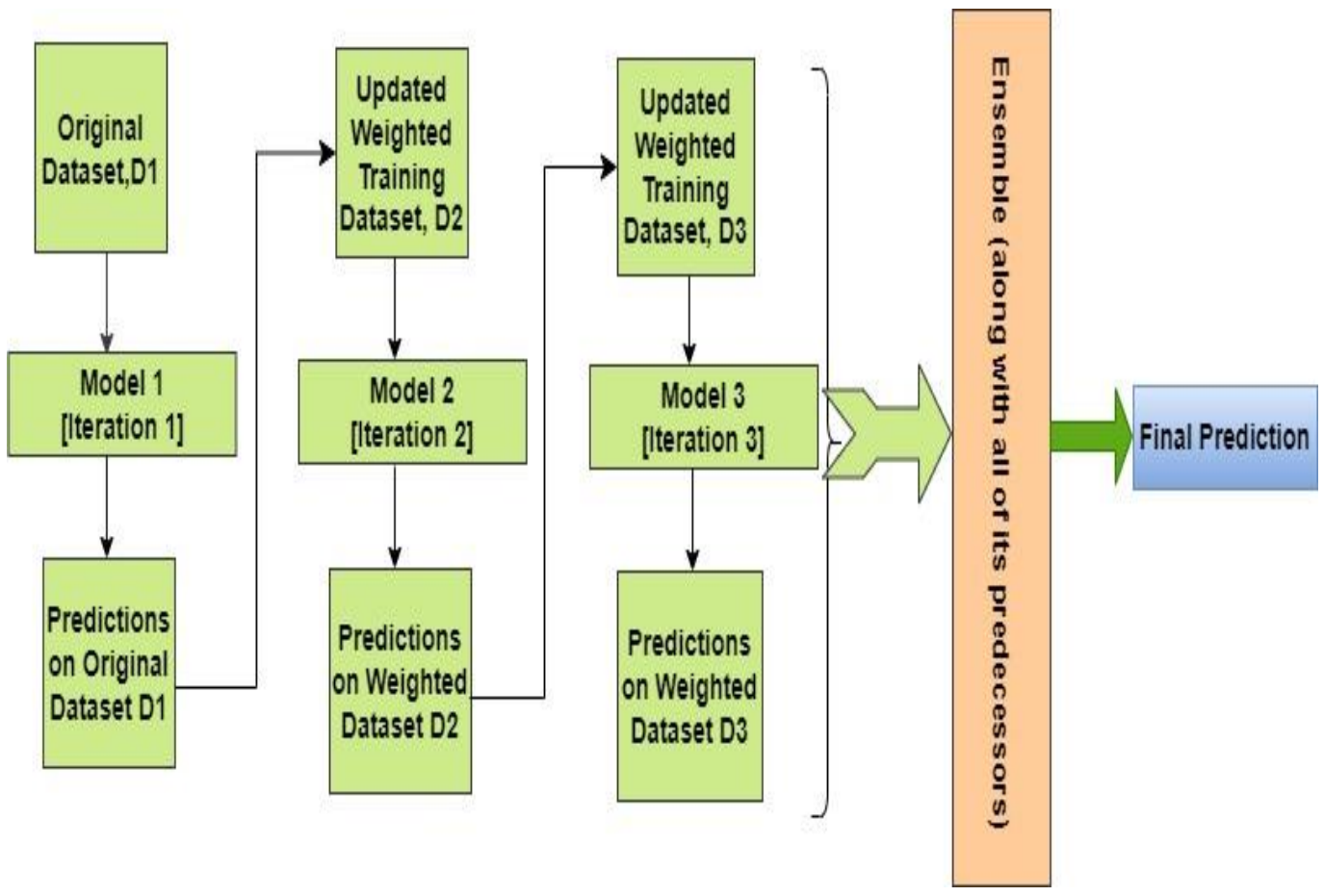

**Figure 23:** The Adaboost Algorithm's structural architecture

### 4.9. Artificial Neural Network (ANN)

Artificial neural networks (ANNs) are efficient problem-solving techniques that mimic the human brain's structure, making them more efficient in tasks like pattern recognition than traditional digital computers. ANNs are parallel processors used for storing experimental information and addressing multivariate and nonlinear modeling issues. They are often used as surrogate or response surface approximation models [57]. ANNs consist of an input, hidden, and output layer with adjustable weights for artificial neurons. The activation functions of artificial neurons range from (-1 to 1), with logarithmic and tangent sigmoids being popular. ANN design involves determining the number of inputs, outputs, hidden layers, and hidden neurons.

The three core layers of an artificial neural network are outlined below:

- **Input layer**: To use the model's neural network, it initializes data. The actual value from the data is found in the input layers. Each input sent by the computer programmer is approved by the input layer, which can be determined by the pre-processed dataset's attributes.
- **Hidden layer**: Everything that has to be calculated is performed in this layer of code, which functions as a bridge between the input and output layers where the extensive network is composed of three or more layers. It executes all the computations essential to reveal hidden patterns and features that contribute to the outcome.
- **Output layer**: The outcome emerges (normal or anomaly according to the assault classifications).

Each of the nodes inside the input layer is entirely linked with every other node within the subsequent hidden layer, and so on throughout all remaining layers. Figure 24, illustrates that the nodes' interconnections can be viewed as a connected graph. The hidden layer is implemented to modify the input through an assortment of operations that eventually generate an output that is communicated via this layer of code.

The artificial neural network calculates the weighted sum of inputs after receiving input using a transfer function to add a bias.

$$Y = \sum_{i=1}^{n} Wi*Xi + b \qquad (5)$$

Where W represents the weight and b signifies bias.
and y is the output of the model.

The weighted total is input to an activation function, which determines whether a node fires or not, reaching the output layer only for activated neural networks.

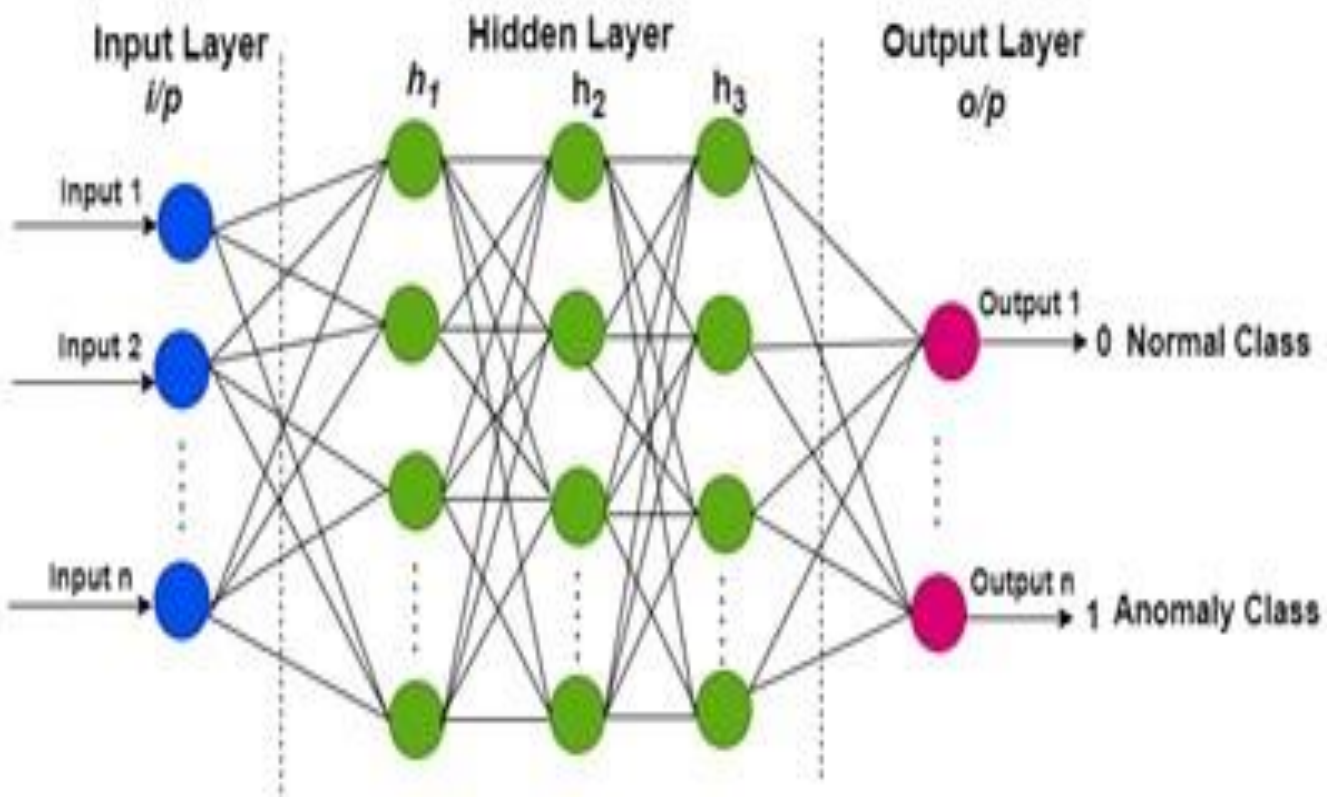

**Figure 24:** Structure of Artificial Neural Network

### 4.10. Deep Neural Network (DNN)

Deep Neural Networks (DNNs) enhance Intrusion Detection Systems (IDS) by identifying complex network data patterns. Data collection, feature extraction, and multiple-layer DNN design are done with hyperparameter optimization and overfitting techniques. The Evaluation focuses on accuracy, recall, and F1-Score. Challenges like data imbalance and high computational demands are addressed. DNNs are deployed for real-time intrusion detection, with regular updates to maintain effectiveness.

Intrusion Detection Systems (IDS) are crucial for network security, identifying and mitigating unauthorized activities. Integrating deep neural networks (DNN) with IDS enhances their effectiveness by recognizing patterns and anomalies in network data. DNNs can adapt to complex environments, improving their accuracy and efficiency in threat detection. This combination offers robust protection against emerging and sophisticated threats.

Deep Neural Networks (DNNs) are a powerful technique for Intrusion Detection Systems (IDS) by learning complex patterns from network traffic data [58]. This is implemented by collecting comprehensive datasets, preprocessing through feature selection, normalization, and augmentation, and training using cross-entropy loss and optimizers. DNNs are tested on unseen data and deployed in real-time environments, improving detection accuracy and capturing intricate attack patterns.

The detection accuracy of intrusion detection, based on the Deep Neural Network (DNN) algorithm model's overall detection level, is the proportion of correctly identified normal and anomaly class and also assists in developing a resilient and efficient intrusion detection system (IDS) that can identify and classify unpredictable and abnormal cyberattacks.

### 5. Table 8: A Ten-Year Comparative Overview of Related Works

| ***Reference*** | ***Dataset*** | ***Classifiers Applied*** | ***Evaluated Matrix With Accuracy*** | ***Findings*** |
|---|---|---|---|---|
| Chakrawarti, A. ., & Shrivastava, S. S.[51] | NSL-KDD and AWID | Deep Q-Networks (DQN) | **Accuracy**<br>Deep Q-Networks (DQN) = 96.8% | "IDS Method Based on Deep Q-Networks"<br>• Achieves 96.8% accuracy.<br>• Showcases elegant techniques for improved precision. |
| Mohammad Almseidin, Maen Alzubi, et al. [59] | KDD CUP 99 | J48, Random Forest, Random Tree, Decision Table, MLP, Naive Bayes, and Bayes Network | **TP, TN, FN, FP, Precision**<br>J48 = 93.10%<br>Random Forest = 93.77%<br>Random Tree= 90.57%<br>Decision Table=92.44%<br>MLP=91.90%<br>Naïve Bayes=91.23%<br>Bayes Network=90.73% | Random Forest Classifier Features<br>• Low RMSE score.<br>• Lowest false-positive rate.<br>• Highest accuracy rate of 93.77%. |
| Manjula C. Belavagi et al. [60] | NSL-KDD | Logistic Regression, Gaussian Naive Bayes, Support Vector Machine, and Random Forest. | **Precision, Recall, F1-Score**<br>LR = 0.84<br>GNB = 0.79<br>SVM = 0.75<br>RFC = 0.99 | Random Forest Classifier's Intrusion Detection Accuracy<br>• Outperforms 99%<br>• Allows further study on key characteristics<br>• Supports multiclass classification classifiers. |
| T.Saranyaa, S.Sridevi et al. [61] | KDD cup 99 | Modified K-means, SVM, J48, Naïve Bayes, Decision Table, PCA-LDA-SVM, Logistic Regression, Decision Tree, ANN, LDA, CART, Random Forest | **Accuracy, Precision, Recall, and F-Score**<br>Modified K-means = 95.75%<br>SVM = 98.9%<br>J.48 = 99.12%<br>Naïve Bayes = 92.7%<br>Decision Table = 99.45%<br>PCA-LDA-SVM = 92.16%<br>Logistic Regression = 98.3%<br>Decision Tree = 99.65%<br>ANN = 99.65%<br>LDA = 98.1%<br>CART = 98%<br>Random Forest = 99.81 % | ANN, Decision Tree, RF Algorithms for Attack Detection<br>• Performance varies by dataset size and application. |
| Kasongo SM, Sun Y [26] | UNSW-NB15 | ANN, kNN, DT, LR and SVM | **Accuracy, Precision, Recall, and F1-Score**<br>ANN = 94.49%<br>KNN = 96.76%<br>DT = 93.65%<br>LR = 93.22%<br>SVM = 70.98% | XGBoost Enhances Binary Classification Scheme<br>• Test accuracy increases from 88.13 to 90.85%. |
| Mahmood RAR, Abdi A et al. [62] | NSL-KDD | Naïve Bayes, KNN,DT and SVM, GA,PSO (Feature Selection) | **TP, TN, FP, FN**<br>**Accuracy, Precision, Recall, and F-Score**<br>Naïve Bayes = 90.13%<br>KNN = 98.89%<br>DT = 99.38%<br>SVM = 93.55% | Decision Tree Classifier Performance<br>• Outperforms other classifiers in accuracy, precision, recall, f-score.<br>• Optimal feature coupling reduces model-building time and data analysis burden. |
| M. Choubisa, R. Doshi et al. [63] | NSL-KDD | Random Forest (Feature Selection) | **Accuracy, DR, FAR, MCC (Matthew's correlation coefficient)**<br>Dos = 99.69%<br>Probe = 99.69%<br>R2L = 99.68%<br>U2R = 99.69% | "Assault Categorization Model Improvement"<br>• Utilized new feature selection method.<br>• Improved using random forest classifier. |
| Dhanabal, L., Shantharajah [6] | NSL-KDD | J48, SVM, Naïve Bayes (Test Accuracy with 6 features) | **J48** / **SVM** / **NaïveBayes**<br>Normal=99.8 / Normal=98.8 / Normal=74.9<br>DoS =99.1 / DoS =98.7 / DoS =75.2<br>Probe= 98.9 / Probe= 91.4 / Probe= 74.1<br>U2R =98.7 / U2R =94.6 / U2R =72.3<br>R2L= 97.9 / R2L= 92.5 / R2L= 70.1 | J48's Network Classification<br>• Outperforms CFS for data set classification.<br>• Demonstrates potential for network classification. |
| Modi, Urvashi & Jain, Anurag [27] | KDD CUP99 | Bayes Net, Naïve Bayes,J48, J48Graft and Random forest | **Precision, Recall-measure, and ROC**<br>Bayes Net = 0.98<br>Naïve Bayes = 0.96<br>J48 = 0.98<br>J48Graft = 0.98<br>Random Forest = 0.98 | "J48, J48 Graft, Random Forest Machine Learning Algorithms: Over 96% Detection Rate" |
| Mohammadi, Sara & Mirvaziri [64] | KDD CUP99 | FGLCC-CFA, FGLCC, CFA, ID3-BA, N-KPCA-GA-SVM, KMSVM | **AR, DR, FPR**<br>**Methods** **DR** **AR** **FPR**<br>FGLCC-CFA 95.23 95.03 1.65<br>FGLCC 91.32 92.59 2.15<br>CFA 92.05 2.83 3.90<br>ID3-BA 91.57 92.59 3.16<br>N-KPCA-GA-SVM 91.21 92.51 2.01<br>KMSVM 88.71 87.01 - | FGLCC-CFA Filter Performance<br>• Outperforms other methods.<br>• Achieves higher AR and DR.<br>• Lower FPR of 1.65%. |
| Verma P, Shadab K, et al. [18] | NSL-KDD | XGBoost and AdaBoost with KMeans Clustering, XGBoost, AdaBoost | **TP, TN, FP, FN**<br>Accuracy, Precision, Recall, and F1-Score<br>XGBoost with KMeans Clustering = 84.25%<br>Adaboost with KMeans Clustering = 82.01%<br>XGBoost = 80.23%<br>Adaboost = 80.73%<br>XGBoost with K-Means Clustering (Proposed) = 84.25% | "Enhanced Network Intrusion Detection Methods"<br>• Demonstrates accuracy.<br>• Uses machine learning for anomaly detection.<br>• Addresses varying probability distributions. |
| W. L. Al-Yaseen, Z. A. Othman et al. [65] | KDD CUP99 | Multi-level hybrid SVM, Extreme Learning Machine (ELM), Modified | **Accuracy, DR, FAR**<br>Modified K-means (Proposed) ACC=95.75%<br>DR = 95.17%<br>FAR = 1.87% | Modified K-means Enhances KDD Training<br>• Reduces 10% KDD dataset to 99.8%.<br>• Creates high-quality SVM and ELM training datasets.<br>• Enhances multi-level model detection accuracy. |
| G. Yedukondalu, G. H. Bindu et al. [36] | NSL-KDD | SVM, ANN | **Accuracy, Precision, Recall, f1-score**<br>SVM = 48%<br>ANN = 97% | SVM vs ANN Intrusion Detection<br>• SVM: 48% accuracy<br>• ANN: 97% accuracy |
| Tuan-Hong Chua and Iftekhar Salam [66] | CIC-IDS2017 and the CSE-CIC-IDS2018 | DT, RF,SVM, NB, ANN,DNN | **Accuracy, Precision, Recall, and F1-Score**<br>**CIC-IDS2017** **CSE-CIC-IDS2018**<br>DT = 0.9959 DT = 0.5942<br>RF = 0.9967 RF = 0.5949<br>SVM = 0.9600 SVM = 0.7559<br>NB = 0.7296 NB = 0.4972<br>ANN = 0.9549 ANN = 0.7000<br>DNN = 0.9735 DNN= 0.6518 | Experiment Findings:<br>• ANN model optimal for frequent infrastructure upgrades and high cyberattack costs.<br>• DT more effective for systems without frequent updates or significant attacks.<br>• DT best for training and categorizing data. |
| Ghose, Dipayan & Partho et al. [29] | NSL-KDD | MLP,LSTM,NB, DT,KNN,RF,SVM | **Accuracy, Precision, Recall, and F1-Score**<br>NB = 75.9%<br>DT = 88.2%<br>KNN = 87.0%<br>RF = 89.6%<br>SVM = 87.6%<br>MLP = 96.89%<br>LSTM = 97.77% | NSL-KDD Dataset Accuracy<br>• LSTM (97.77%) and MLP (96.89%) used.<br>• Two labels, 41 traffic input features per record. |
| Thaseen, I. S., et al. [67] | NSL-KDD | SVM (Multiclass), Chi-square | **TP and FP Rate, Precision, Recall, F-Measure, ROC area**<br>SVM (Multiclass) = 98% | Experiment on Intrusion Detection<br>• Model achieved 98% accuracy.<br>• Utilizes SVM and chi-square feature selection. |
| K. Dinesh and D. Kalaivani et al. [30] | NSL-KDD | RF, CART(Classification and Regression Trees), SVM, and MLP with different meta-heuristic algorithms (GS, GBO, SA, and GA) | **Accuracy, Precision, Recall, and F1-Score**<br>CART = 89%<br>RF = 93%<br>SVM = 94%<br>MLP = 96% | Intrusion Detection Systems Evaluation<br>• Utilized meta-heuristic and machine learning.<br>• Improved recall, accuracy, precision.<br>• GA-optimized MLP classifier achieved 96% accuracy. |
| Safura A. Mashayak, et.al[68] | NSL-KDD | RF, DT | **Accuracy, Precision, Recall, and F1-Score**<br>**Accuracy**<br>RF = 99.2%<br>DT = 99% | Model's Tolerance Expansion<br>• Model's capabilities expanded to 13 class categorizations.<br>• Despite additional assault classes, model performs exceptionally well. |
| J. Ren, J. Guo, W. Qian, et al. [31] | UNSW-NB15 | Genetic algorithm and Random Forest based feature selection, DO_IDS | **Accuracy, FAR, Macros precision, Macros recall, Macros f1-score**<br>Accuracy<br>RF = 86%<br>DO_IDS = 92.8% | DO IDS: RF Classifier-Based Data Optimization<br>• Outperforms RF classifier in all indicators.<br>• Scores 92.8%. |
| K. -A. Tait et al. [32] | UNSW-NB 15 and CICIDS2017 | RF, KNN, SVM, Binary, and Multiclass | **Accuracy, Precision, Recall, f1-score**<br>**Accuracy**<br>**Binary Class** **Multiclass**<br>RF = 0.9977 RF = 0.9294<br>SVM SVM<br>(Medium Gaussian) =0.9962, Medium Gaussian) = 0.9338, KNN – Weighted = 0.9983 | Experimental Results: Multiclass Classification Enhances Intrusion Detection<br>• Enables targeted attack response. |
| Brao, Bobba et al. [33] | NSL-KDD | K-Nearest Neighbor (KNN) | **Accuracy**<br>KNN = 99.95% | Innovative Approach: 99.6% Accurate Results<br>• Boosts efficiency |
| Nabila Farnaaz and M.A Jabbar [34] | NSL-KDD | Random forest (RF) based ensemble classifier | **Accuracy, DR, FAR, MCC**<br>Accuracy<br>RF = 99.67% | Model's Effectiveness Experiments:<br>• 99.67% scoring system<br>• Low false alarm rate<br>• High detection rate. |
| Lin, Wei-Chao & Ke, Shih-Wen et al. [35] | KDD CUP99 | k-Nearest Neighbor (k-NN), Cluster Center and Nearest Neighbor (CANN), Support Vector Machine (SVM) | **Accuracy**<br>CANN=99.76%<br>KNN=93.87%<br>SVM=80.65% | Feature Encoding Implementation<br>• Implemented for assaults and conventional connectivity.<br>• CANN outperforms 99.76% accuracy.<br>• Compatibility with k-NN and support vector machines. |
| Bhavani T. T, Kameswara M. R et al. [69] | NSL-KDD | Random Forest (RF), Decision Tree (DT) | **Accuracy**<br>RF = 95.323%<br>DT = 81.868% | Random Classifier Performance<br>• Best result: 95.323% success rate.<br>• Easy implementation. |
| Ponthapalli R. et al. [70] | NSL-KDD | Decision Tree (DT), Logistic Regression (LR), Random Forest (RF), Support Vector Machine (SVM) | **Accuracy**<br>RF=73.784%<br>DT=72=303%<br>SVM=71.779%<br>R=68.674% | Research Findings:<br>• Random Forest classifier: Most effective with maximum accuracy of 73.784%. |
| Dutt I. et al. [19] | KDD CUP99 | Feature Selection using Chi-Square Analysis, Feature extraction Frequency episode extraction | **Accuracy**<br>Feature Selection using Chi-Square Analysis = 92.65% | Experiment Results:<br>• Daily improvement in true positive rate accuracy.<br>• Sharp drop in false negative percentage. |
| Maniriho et al.[21] | NSL-KDD, UNSWNB-15 | Single Machine Learning Classifier (K- Nearest Neighbor (KNN)), Ensemble Technique (Random Committee (RC)) | **TPR, FPR, Accuracy, Precision, Mis (Misclassification rate)**<br>NSL-KDD using KNN=98.727%<br>NSL-KDD using RC=99.696%<br>UNSW NB-15 using KNN=97.3346%<br>UNSW NB-15 using RC=98.955% | Ensemble Technique Outperforms Single Classifiers<br>• Model assessed using two distinct datasets. |
| Kazi A., Billal M et al. [71] | NSL-KDD | Artificial Neural Network (ANN), Support Vector Machine (SVM) | **Accuracy**<br>ANN = 94.02%<br>SVM = 82.34% | ANN-based Machine Learning Outperforms SVM in Network Traffic Classification. |
| A. Aziz, Amira & Hanafi et al. [72] | NSL-KDD | Breadth-Forest Tree (BFTree), Naïve Bayes Decision Tree (NBTree), J48, Random Forest Tree (RFT),Multi-Layer Perceptron (MLP),Naïve Bayes | **TP, FP, FN, Precision, Recall, F-Score**<br>**Accuracy**<br>BFTree=98.24%<br>NBTree=98.44%<br>J48=97.68%<br>RFT=98.34%<br>MLP=98.53%<br>NB=84.75% | Improved False Positives Percentage<br>• NBTree and BFTree outperformed J48 and RFTree.<br>• MLP scored highest in DoS and Normal classifications.<br>• Struggled with R2L and U2R assaults. |
| Alkasassbeh, M & Almseidin, M [73] | KDD CUP'99 | J48 Tress, Multilayer Perceptron (MLP), Bayes Network | **TP Rate FP Rate Precision ROC Area**<br>Accuracy<br>J48=93.1083%<br>MLP=91.9017%<br>Bayes Network=90.7317% | J48 Classifier's Accuracy Improvement<br>• Solved low assault detection issue.<br>• Achieved highest accuracy rate for KDD dataset attacks |
| Rasane, Komal & Bewoor et al. [74] | KDD-Cup 99, NSL-KDD, UNSW-NB 15, Kyoto 2006+ | KNN.NB, SGD, DT, RF | **Accuracy**<br>KNN = 97.8%<br>NB = 89%<br>SGD = 94.9%<br>DT = 97.9%<br>RF = 98% | "RF Algorithm Outperforms Other Machine Learning Methods for Data Categorization"<br>• Experimental results show superior performance on RF. |
| Anwer, H.M., Farouk, M., et al. [75] | UNSWNB-15 | J48 and NB | **Accuracy**<br>J48 = 88%<br>NB = 76% | Experimental Results:<br>• Use 18 GR ranking features and J48 classifier.<br>• Achieve 88% accuracy rate. |
| Ravale, Ujwala & Marathe et al. [76] | KDD CUP'99 | Hybrid k-means and SVM-RBF | **Accuracy**<br>KMSVM = 88.71% | KMSVM Algorithm Outperforms KM and SVM<br>• Improves accuracy results. |
| Khammassi, Chaouki et al. [47] | KDDCup 99, UNSW-NB15 | Genetic Algorithm (GA) as search and Logistic Regression (LR) as learning algorithm | **Accuracy, FAR**<br>KDD CUP'99 (GALR-DT) = 99.90%<br>UNSW-NB15 (GALR-DT) = 81.42% | Experiment Results:<br>• High classification accuracy with 18 characteristics.<br>• 99.90% DR, 99.81% DR, 0.105% FAR.<br>• Utilized KDD99 dataset. |
| Kotpalliwar MV et al. [77] | KDD CUP'99 | SVM | **Accuracy**<br>Validation Accuracy = 89.85%<br>Classification Accuracy = 99.9% | "KDD Dataset Analysis"<br>• 10% datasets varied in assault types and samples.<br>• Resulted in "mixed" dataset with 99.9% accuracy. |
| A, Anish Halimaa; Sundarakantham, K. [78] | NSL KDD | SVM, Naïve Bayes | **Accuracy, Misclassification Rate**<br>SVM = 97.29%<br>Naïve Bayes = 67.26% | "SVM Outperforms Naïve Bayes in Machine Learning"<br>• Higher accuracy rate (97.29)<br>• Lower misclassification rate (2.705) |
| Basheri, Mohammad & Iqbal et al. [22] | NSL KDD | SVM, RF, ELM | **Accuracy, Recall**<br>SVM (Linear) = 99.2%<br>RF = 97.7%<br>ELM = 99.5% | NSL KDD Dataset for Intrusion Detection<br>• Utilized for knowledge discovery and data mining.<br>• ELM outperforms other strategies. |
| S. Teng, N. Wu, H. Zhu et al [80] | KDD CUP'99 | Single Type-SVM, CAIDM | **Accuracy**<br>Single Type-SVM = 81.72%<br>CAIDM = 89.02% | Optimized CAIDM:<br>• Based on 2-class SVMs and DTs.<br>• Enhances accuracy and efficiency. |
| B. S. Bhati and C. S. Rai [49] | NSL-KDD | SVM | **Accuracy, Error Rate Error Rate**<br>Linear SVM = 96.1% 3.9%<br>Quadratic SVM = 98.6% 1.4%<br>Fine Gaussian SVM = 98.7% 1.3%<br>Medium Gaussian SVM = 98.5% 1.5% | Analysis of SVM Detection Accuracy<br>• Linear SVM, quadratic SVM, fine Gaussian SVM.<br>• Medium Gaussian SVM has varying accuracy.<br>• Fine Gaussian SVM offers best accuracy and minimal error. |
| D. Gupta, S. Singhal, et al. [81] | NSL-KDD | Data mining techniques, Linear Regression, K-Means Clustering | **Accuracy**<br>Linear Regression = 80%<br>K-Means Clustering = 67.5% | Network Assault Identification<br>• Linear regression: 80% accuracy<br>• K-means clustering algorithm: 67.5% accuracy. |
| K. Goeschel [82] | KDDCup '99 | SVM, DT, Naïve Bayes techniques | **Overall Accuracy** = 99.62% | "Accuracy in Final Phase: 99.62%"<br>• False Positive Rate: 1.57%<br>• Higher FPR: 4.29%. |
| Iqbal, A., & Aftab, S.[46] | NSL-KDD | Forward Neural Network (FFANN), Pattern Recognition Neural Network (PRANN) | **Accuracy, MCC, R-squared, MSE, DR, FAR and AROC**<br>FFANN=98.0792%<br>PRANN=96.6225% | "Multiple Classifier Combination Enhances Performance"<br>• Both models show superior performance in attack detection metrics. |
| Shyla,kapil kumar et al. [83] | KDD Cup99 | Naïve Bayes, Linear SVM, Random Forest | **Accuracy, Precision, Recall and F1-Score**<br>Naïve Bayes = 0.971<br>SVM = 0.994<br>Random Forest = 0.999 | KDD Cup99 Dataset Comparison<br>• Compared algorithms' accuracy, precision, detection rate.<br>• Random Forest ranked highest with 0.999 detection rate. |
| Deyban P. Miguel A. A [20] | NSL-KDD | A hybrid model of supervised (Neural Network (NN), Support Vector Machine (SVM)), and unsupervised (K-Means) machine learning algorithms. | **Accuracy, Error rate, Sensibility, Specificity, Precision, ROC Curve**<br>SVM+K-Means=96.81%<br>NN+K-Means=95.55% | Enhancing IDS Effectiveness<br>• Incorporating supervised and unsupervised learning methods. |
| Aburomman, Abdulla & Reaz et al. [84] | KDDCup '99 | PCA-SVM,LDA-SVM,PCA-LDA-SVM | **Accuracy**<br>PCA-SVM = 0.8902<br>LDA-SVM = 0.8993<br>PCA-LDA-SVM = 0.9216<br>Overall-accuracy (ACC) = 0.92162, False-positive (FP) = 0.0196, False-negative (FN) = 0.10849 | Ensemble PCA-LDA-SVM Method: Outperforms Single Feature Extraction<br>• Improves accuracy and performance. |
| Al-Jarrah, O. Y., Al-Hammdi et al. [85] | NSL, Kyoto 2006+ | SMLC,AdaBoost M1,Bagging,RF | **Accuracy**<br>**NSL KDD** **Kyoto 2006+**<br>SMLC = 99.58% SMLC = 99.39%<br>AdaBoostM1 = 94.20% AdaBoostM1= 95.88%<br>Bagging = 99.55% Bagging = 99.39%<br>RF = 99.62% RF = 99.37% | SMLC Outperforms Supervised Ensemble ML on Network Intrusion Datasets<br>• Detection accuracy comparable to supervised ensemble models.<br>• 20% fewer labeled training data instances. |
| B. M. Irfan, V. Poornima et al. [86] | KDD CUP'99, NSL KDD | DT,RF,SVM,NN, DL models | **Accuracy, Precision, Recall, FPR. F1-Score**<br>KDD CUP'99 NSL-KDD<br>DT = 0.85 DT = 0.82<br>RF = 0.89 RF = 0.85<br>SVM = 0.87 SVM = 0.84<br>NN = 0.88 NN = 0.86<br>DL models = 0.90 DL models = 0.88 | Deep Learning Models for Intrusion Detection<br>• Convolutional and recurrent neural networks enhance detection.<br>• Random forests improve recall. |
| M. D. Rokade and Y. K. Sharma [87] | NSL-KDD | Naïve Bayes, SVM, ANN, RF | **Accuracy, Precision, Recall, f1-score**<br>NB = 98%<br>SVM = 95%<br>ANN = 95%<br>RF = 88% | Experimental Study on Anomaly Detection<br>• Uses SVM, Naïve Bayes, ANN.<br>• Demonstrates real-time network performance. |
| S. Waskle, L. Parashar et al. [88] | KDD CUP'99 | SVM, Naïve Bayes, DT, PCA With Random Forest | **Accuracy, Error Rate**<br>SVM = 84.34%<br>NB = 80.85%<br>DT = 89.91%<br>PCA With RF = 96.78% | Proposed Technique Outperforms SVM, Naive Bayes, Decision Trees<br>• 96.78% accuracy rate<br>• Minimal 3.24 minute performance time. |
| M. Hammad, W. El-medany et al. [38] | UNSWNB-15 | Naïve Bayes, J48, RF, ZeroR | **Accuracy, Precision, Recall, f1-score, FPR, Specificity**<br>NB = 76.04%<br>J48 = 93.78%<br>RF = 97.60%<br>ZeroR = 68.06% | J48 and RF Algorithms: Favorable Outcomes. |
| A. Singhal, A. Maan et al. [39] | NSL-KDD | KNN, DT, NB, SVM | **Accuracy**<br>**Independent Accuracy Inference Function.**<br>KNN = 95.755% KNN = 97.605%<br>DT = 98.088% DT = 98.554%<br>NB = 82.971% NB = 66.687%<br>SVM = 81.263% SVM = 93.914% | "Inference Detection Model Enhances Factors"<br>• Enhances detection of unobservable factors.<br>• Uses SVM technique for improved accuracy. |
| J. D. S. W.S. and P. B. [40] | KDD CUP'99 | LDA, SVMr, RF, ADABoost | **Sensitivity(TPR) Specificity (TNR) ROC**<br>LDA = 0.9762 0.7121 0.8827<br>SVMr = 0.9933 0.9511 0.9782<br>RF = 0.9884 0.9495 0.9718<br>ADABoost = 0.9975 0.9569 0.9824 | "Inference Detection Model Enhances Factors"<br>• Enhances detection of unobservable factors.<br>• Uses SVM technique for improved accuracy. |
| P. V. Pandit, S. Bhushan, et al. [43] | CICIDS2017 | RF,XG Boost, Extra Tree,DT | **Accuracy, Precision, Recall, f1-score,**<br>RF = 0.9899<br>XG Boost = 0.9929<br>Extra Tree = 0.9935<br>DT = 0.9946 | Python Technique Enhances Detection Accuracy<br>• Utilizes precision, recall, f1-score metrics.<br>• Outperforms state-of-the-art technology. |
| Anouar Bachar, N. E. [44] | UNSWNB-15 | SVM model for binary classification | **TPR, FPR, Accuracy, Precision, Recall, f1-score**<br>SVM Polynomial = 94%<br>SVM Gaussian = 93% | Simulation Results:<br>• SVM-Gaussian model improves accuracy by 93%<br>• SVM-Polynomial model enhances accuracy by 94%. |
| Nitu Dash, S. C. [45] | NSL KDD | MLP, Radial Basis Classifier, and gradient descent Backpropagation learning algorithm | **Accuracy**<br>PCA + MLP = 97.8803 %<br>PCA + RBF Classifier = 98.1162% | PCA Feature Extraction Model<br>• Enhances accuracy in short computational time.<br>• RBF: 98.1162% accuracy in 36.28 seconds.<br>• MLP: 97.8803% accuracy in 41.37 seconds. |
| Kumar, G. K. [46] | UNSWNB-15 | KNN, LR, NB, SGD and RF | **Accuracy, Precision, Recall, F1-Score, MSE, TPR, FPR**<br>KNN = 98.28%<br>LR = 98.42%<br>NB = 76.59%<br>SGD = 98.16%<br>RF = 99.57% | RF Classifier Performance on UNSWNB-15<br>• Outperforms other classifiers.<br>• Accuracy: 99.57% with all features, 99.64% with some features. |
| Senthilnayaki, B., Venkatalakshmi [50] | KDD CUP'99 | Modified-KNN,SVM and KNN | **Detected assault accuracy**<br>M-KNN = 98.58% | Modified KNN Feature Selection<br>• Reduces undesirable traits.<br>• Improves security.<br>• Reduces false alarm rates.<br>• Outperforms other algorithms. |
| J. Gao, S. Chai, C. Zhang et al. [89] | UNSWNB-15 | ELM, MVT (Multi-Voting Technology) | **Accuracy, TP, TN, FP, FN**<br>Accuracy = 89.71% | "MVT Improves IDS Accuracy"<br>• Superior to IDS without MVT.<br>• Suggests strategy for high detection accuracy.<br>• Reduces time required. |
| M. Zaman and C.-H. Lung [90] | Kyoto 2006+ | KM,KNN,FCM,NB,SVM, RBF, Ensemble | **Precision, Recall, Accuracy, ROC**<br>**Accuracy** **ROC**<br>KM = 0.836 KM = 0.6148<br>KNN = 0.9754 KNN = 0.9532<br>FCM = 0.836 FCM = 0.6148<br>NB = 0.9672 NB = 0.9481<br>SVM = 0.9426 SVM = 0.8023<br>RBF = 0.9754 RBF = 0.9741<br>Ensemble = 0.9672 Ensemble = 0.9639 | RBF Classification Method Performance<br>• Superior accuracy: 0.9754<br>• Outperforms Ensemble approach: 0.9631 |
| Mazarbhuiya, Fokrul et al. [91] | KDDCUP'99, Kitsune | IFRSCAD (Intuitionistic Fuzzy-Rough Set-Based Classification for Anomaly Detection) | **Normal TPR, Attack TPR**<br>Normal detected assaults<br>KDD CUP'99<br>IFRSCAD (Normal TPR) = 96.99% and 91.289%<br>Kitsune<br>IFRSCAD (Attack TPR) = 96.29 and 91.289% | "Proposed Algorithm Outperforms Classification-Based Algorithms"<br>• Extracts anomalies with 96.99% accuracy.<br>• Demonstrates superior performance with KDDCUP'99 and Kitsune datasets. |
| Shen Kejia, Hamid Parvin et al [92] | NSL-KDD | FRSTSS, FRSTFS (Fuzzy Rough Set Theory based Sample Selection and Feature Selection ), SVM | **Accuracy of detected assaults**<br>FRSTSS+FRSTFS+SVM = R2L, U2R, Probe, DoS, Normal = 99.21%, 57.07%, 99.94%, 97.98%, 97.02% | Network-Based Intrusion Detection Systems Evaluation<br>• Uses fuzzy rough set theory and SVM.<br>• Highlights feature selection techniques' efficiency. |
| Sever, Hayri & Raoof et al. [93] | ADFA-LD and NSL-KDD | RSC(Rough Set Classification) approach using MODLEM algorithm, KNN, SVM, NB, DT | **Precision, Recall, F-Score**<br>RSC = 88.9<br>KNN = 86.2<br>SVM = 82.0<br>NB = 74.8<br>DT = 86.1 | Study on Attribute Reduction Strategy<br>• Demonstrates impact on classification accuracy.<br>• RSC model achieves 88.9% F-score.<br>• Uses fuzzy, rough set of ten features. |
| Q. Zhang, Y. Qu et al. [94] | KDD Cup 99 | MFNN ( Multi-functional nearest-neighbour), NB, SMO, IBK (Instance-based), Ada-boost and RF | **Detected assaults**<br>MFNN = 99.62%<br>NB = 91.03%<br>SMO = 97.30%<br>IBk = 99.64%<br>Ada-Boost = 99.89%<br>RF = 99.93% | "Evaluating Feature Selection Techniques"<br>• Focuses on KFRFS for accuracy and reduction.<br>• Highlights superior computational efficiency. |
| Panigrahi, Ashalata & Patra, Manas [95] | NSL-KDD | Fuzzy NN, Fuzzy-Rough NN, FRONN, VQNN, OWANN. | **Accuracy, Precision, Recall, FAR**<br>Random Search<br>Fuzzy NN = 94.5591<br>Fuzzy Rough NN = 99.5951<br>VQNN = 99.3991<br>Fuzzy Ownership NN = 99.6086<br>OWANN = 99.388 | Study on Classifier Performance<br>• Evaluates accuracy, detection rate, precision, false alarm rate.<br>• Finds fuzzy ownership nearest neighbor classification with random search superior. |
| Tripathy, S. S., & Behera, B. [4] | KDD Cup 99 | KNN, DT, MNB, BNB, RF, SVM, PPN, LR, XGBOOST, AdaBoost, SGD, Ridge, RC, PA, BPN | **Accuracy, Precision, Recall, F-1 Score**<br>KNN = 0.9724, DT = 0.9713, MNB = 0.9329, BNB = 0.9473, RF = 0.9714, SVM = 0.9808, PPN = 0.9101, LR = 0.9667, XGBOOST = 0.9464, AdaBoost = 0.9431, SGD = 0.9046, Ridge = 0.9495, RC = 0.9487, PA = 0.9443, BPN = 0.9704 | SVM Classifier Performance<br>• Outperforms other classifiers with 98.08% score.<br>• DT, RF, BPN, KNN perform well. |
| Kushal Jani, Punit Lalwani et al. [96] | NSL-KDD | KNN,RF,DT,NB, SVM,ANN,DNN | **Accuracy, Precision, Recall**<br>KNN = 76.47%<br>RF = 78.43%<br>DT = 80.95%<br>NB = 82.07%<br>SVM = 82.63%<br>ANN = 84.87%<br>DNN = 86.75% | Deep Learning vs Artificial Neural Networks<br>• Deep learning: 86.75% accuracy<br>• Artificial Neural Networks: 84.87% accuracy<br>• Minimal false negatives and positives. |
| Kocher, Geeta & Kumar Ahuja [97] | UNSW-NB15 | KNN,NB,RF,SG D,LR | **Accuracy, Precision, Recall, F-1 Score**<br>LR = 98.17%<br>NB = 75.16%<br>RF = 99.64%<br>SGD = 97.99%<br>KNN = 98.90% | RF Classifier Performance on UNSW15 Dataset<br>• Outperforms other classifiers with 99.64% accuracy.<br>• Potential for multiclass classification intrusion detection. |

## 6. Conclusions and Future Works

Fast growth in the web, including networking sites and current communication methods, led to a boom of networking information and, finally, with it, an unprecedented variety of new hazards to computer security. More investigators are now developing and putting into practice more complex strategies, applying machine learning (ML) techniques like SVM, KNN, DT, LR, NB, RF, XGBOOST, Adaboost, and ANN, to help combat these threats to our digital lives in response to the aforementioned latest developments. Among these instances, SVMs are regarded as one of the distinctive algorithms used in machine learning for intrusion detection, solely owing to their exceptional extrapolation authority across dataset sizes and their capacity to avert the curse of dimensionality. We have discussed IDS datasets and their specifications in this review paper, including prior research on the various IDS types and various machine learning (ML) classification algorithms. Along with presenting an in-depth review of the various datasets, such as KDDCUP'99, NSL-KDD, UNSW-NB15, CICIDS-2017, and CSE-CIC-IDS2018, we additionally discussed the significance of using different ML classifiers in the detection of intrusions and performance evaluation.

We have reviewed research studies that implement the machine learning (ML) classification algorithms pointed out above for the detection of intrusions, counting their techniques and methods of operation. Furthermore, we have offered a tabulated summary and critical analysis of each of these methods, which were assessed using five different datasets and ML classifiers. This highlights the functions of the various algorithms that were employed, as well as the attacks that were discovered and the results of the performance evaluation, including accuracy metrics.

Advancements in machine learning techniques are being used by academics and researchers to develop classification models for intrusion detection systems. This research paper reviews studies from the past decade that have implemented machine learning classification methods to enhance the performance of intrusion detection systems with high accuracy and low false alarm rates.

We plan to extend this research in the future by incorporating a systematic analysis and review of IDS applied to other popular digital security or industry-focused real-time intrusive datasets for machine and deep learning methodologies, like CNN, RNN, deep autoencoders, and generative adversarial networks (GANs).

These above methods have significant potential in computer networking fields, potentially enhancing the reliability and efficiency of intrusion detection systems.

**Acknowledgements**

My sincere gratitude goes out to my mentor, Dr. Bichitrananda Behera of the Department of Computer Science and Engineering at C.V. Raman Global University in Bhubaneswar, Odisha, who helped me finish this topic and guided me in pursuing it. Additionally, I want to express my gratitude to my family and brother for his amazing advice and unceasing support. I consider it an honor that I was able to conduct my research with his help.

**Author contributions:**

1. **Sudhanshu Sekhar Tripathy:** Conceptualization, Algorithm design, Literature study, Dataset analysis

2. **Dr. Bichitrananda Behera**: Algorithm validation, Dataset Investigation

**Conflict of Interest**

The authors declare that they have no conflict of interest.

**Competing Interests**

The authors have no competing interests to declare that are relevant to the content of this article.

**Funding Details**

No funding was received to assist with the preparation of this manuscript.